\documentclass[review,latin9,utf8]{elsarticle}

\usepackage[T1]{fontenc}
\usepackage{color}

\usepackage{graphicx}
\usepackage[skip=0.5\baselineskip]{caption}

\usepackage{esint}
\usepackage{lineno,hyperref}
\usepackage[T1]{fontenc}

\modulolinenumbers[100]

\providecommand{\tabularnewline}{\\}

\makeatother

\begin{document}

\begin{frontmatter}

\title{Quark-Meson-Coupling (QMC) model for finite nuclei, nuclear matter and beyond}

\author{P. A. M. Guichon}
\address{IRFU-CEA, Universit\'e Paris-Saclay, F91191 Gif sur Yvette, France}

\author{J. R. Stone\fnref{myfootnote}}
\address{Department of Physics (Astro), University of Oxford, OX1 3RH United Kingdom}
\address{Department of Physics and Astronomy, University of Tennessee, TN 37996 USA}
\fntext[myfootnote]{Corresponding author email: jirina.stone@physics.ox.ac.uk}

\author{A. W. Thomas}
\address{CSSM and CoEPP, Department of Physics, University of Adelaide, SA 5005 Australia}

\begin{abstract}
The Quark-Meson-Coupling model, which self-consistently relates the dynamics of the internal quark structure of a hadron to the relativistic mean fields arising in nuclear matter, provides a natural explanation to many open questions in low energy nuclear physics, including the origin of many-body nuclear forces and their saturation, the spin-orbit interaction and properties of hadronic  matter at a wide range of densities up to those occurring in the cores of neutron stars. Here we focus on four aspects of the model (i) a full comprehensive survey of the theory, including the latest developments,  (ii) extensive application of the model  to ground state properties of finite nuclei and hypernuclei, with a discussion of similarities and differences between the QMC and Skyrme energy density functionals, (iii) equilibrium conditions and composition of hadronic matter in cold and warm neutron stars and their comparison with the outcome of relativistic mean-field theories and, (iv) tests of the fundamental idea that hadron structure changes in-medium. 
\end{abstract}


\end{frontmatter}
\tableofcontents{}
\linenumbers
\section{Introduction}
\label{intro}
Even though it is a little over a century since the discovery of the atomic nucleus there are still many mysteries to be unravelled. In the beginning, of course, it was impossible to imagine how to build such a small object from protons and electrons, the only elementary particles known at that time. With the discovery of the neutron one had a clear path forward and much of nuclear theory has been concerned with the solution of the non-relativistic many body problem with neutrons and protons interacting through two- and three-body forces. Traditionally this involved the development of phenomenological nucleon-nucleon potentials fit to world data. Early work was based on the one-boson exchange model, followed by so-called ''realistic forces'', which were typically local. Three-body forces initially involved the excitation of an intermediate  $\Delta$ resonance. In recent years a great deal of interesting work has been carried out within the formalism of chiral effective field theory, in which the relevant degrees of freedom are taken to be nucleons and pions. All of these approaches typically involve 20-30 parameters to describe the NN force, with up to five or six additional parameters tuned to some nuclear data in order to describe the three-body force.

Only in the 1970s were quarks discovered. As the fundamental degrees of freedom for the strong force, it is natural to ask whether or not they play a role in the structure of the atomic nucleus. For many the simple answer is no and at first glance the argument seems sound. The energy scale for exciting a nucleon is several hundred MeV, while the energy scale for nuclear binding is of order 10 MeV and this certainly suggests that one should be able to treat the bound nucleons as essentially structureless objects. However, after a little reflection one may be led to at least consider the possibility that  changes in the internal structure of bound nucleons might be relevant. The key ideas are the following:\\
\begin{itemize}
\item Relativity

Since the 1970s the study of the NN force using dispersion relations, primarily by the Paris group of Vinh Mau and collaborators and the Stony Brook group of Brown and collaborators, established in a model independent way that the intermediate range NN force is an attractive Lorentz scalar. On the other hand, the shorter range repulsion has a Lorentz vector character.
\item The Lorentz scalar mean field is large

Once one realizes the different Lorentz structure of the various components of the nuclear force it becomes clear that the very small average binding of atomic nuclei is the result of a remarkable cancellation between large numbers. Whether one starts from the original ideas of a one-boson exchange force from the 1960s, where the scalar attraction was represented by $\sigma$ meson exchange and the vector repulsion by $\omega$ meson exchange, or the related treatment within Quantum Hadrodynamics (QHD) in the 1970s and 80s, or even if one looks at modern relativistic Brueckner-Hartree-Fock calculations, it is clear that the mean scalar attraction felt by a nucleon in a nuclear medium is  several hundreds of MeV. Indeed, within QHD this scalar field was of the order of half of the nucleon mass at nuclear matter density.
\item The Effect on Internal Hadron Structure Depends on the Lorentz Structure

Whereas the time component of a Lorentz vector mean field  simply rescales energies, an applied scalar field changes the dynamics of the system, modifying the mass of the quarks inside a bound hadron. The former makes no change to the internal structure whereas the latter can, in principle, lead to significant dynamical effects.
\end{itemize}

A priori then, the naive argument concerning energy scales is clearly incorrect and it is a quantitative question of whether or not the application of a scalar field, with a strength up to one half of the mass of the nucleon, actually does lead to any significant change in its structure. The only way to answer this question at present is to construct a model of hadron structure, couple the quarks to the relativistic mean fields expected to arise in nuclear matter and see what happens. This was the approach taken by Guichon in 1988 \cite{Guichon1988}, with what has become known as the quark-meson coupling (QMC) model. There the effect of the scalar field was computed using the MIT bag model to describe the light quark confinement in a nucleon. In that model it was found that indeed the applied scalar field {\em did} lead to significant changes in the structure of the bound nucleon. 

Of course, in the context of the famous European Muon Collaboration (EMC) effect, which had been discovered just a few years before, some very dramatic changes to the structure of a bound nucleon had already been proposed. Several of those proposed explanations involved a dramatic ''swelling'' of the bound nucleon, by as much as to 20\% at nuclear matter density, or even the appearance of multi-quark states. But those proposals, by and large, came from the particle physics community and nuclear theory tended to ignore them. Indeed, the EMC effect is still hardly mentioned in the context of nuclear structure; the one exception being the proposal that the structure of the relatively small number of nucleons involved in short-range correlations might be dramatically modified, while the rest remain immutable.

The QMC approach of applying the scalar mean field $\sigma$ self-consistently to a bound nucleon and calculating the consequences was less spectacular than the speculations inspired by the EMC effect, at least at a superficial glance. In particular, the size of the bound nucleon did not increase dramatically (e.g., an increase of only 1-2\% in the confinement radius). On the other hand, the lower Dirac components of the valence quark wave functions (see Section~\ref{BaryonInExternal} for details) were significantly enhanced -- a very natural result given the light mass of the $u$ and $d$ quarks. Since the scalar coupling to the nucleon goes like the integral over the {\em difference} of squares of the upper and lower components, this means that the effective scalar coupling constant for the $\sigma$ to the composite nucleon decreases with density. By analogy with the well-known electric and magnetic polarizabilities, which describe the response of a nucleon to applied electric and magnetic fields, this effect was parameterized in terms of a ''scalar polarizability'', $d$. The nucleon effective mass, calculated self-consistently in an applied scalar field  $\sigma$ of strength, $g_\sigma\sigma$, was then (to a good approximation)
\begin{equation}
M_N^* = M_N - g_\sigma\sigma +\frac{d}{2} (g_\sigma\sigma)^2 \, .
\label{eq:Mstar}
\end{equation}
In Eq.~(\ref{eq:Mstar}) $g_\sigma$ is the coupling of the scalar meson, which represents the attractive Lorentz force between nucleons, to the composite nucleon {\em in free space}. For the MIT bag model, the scalar polarizability was $d \approx 0.18 R_B$, with $R_B$ the bag radius.

While the significance of this reduction of the strength of the $\sigma$ coupling to the nucleon with increasing density may not be immediately apparent, it is extremely important. For example, it has the effect of leading to the saturation of nuclear matter far more effectively than in QHD, where one has to have very large scalar fields in order to reduce the nucleon scalar density sufficiently far below the time component of the nucleon vector density (i.e., until $\bar{\psi} \psi$ is significantly smaller than $\psi^\dagger \psi$, with $\psi$ being the nucleon field), which is what leads to saturation there. As a result, the mean scalar field is considerably smaller in the QMC model than in QHD. In more recent developments of the model, where an equivalent energy density functional was derived which could be used to calculate the properties of finite nuclei, the scalar polarizability naturally generates a three-body force.

Because the QMC model has as input the coupling of the $\sigma, \omega, \rho$ and $\pi$ mesons to the {\em light quarks}, one can calculate the binding of any hadron in nuclear matter, {\em with no new parameters}, provided one ignores any OZI suppressed coupling of these mesons to heavier quarks. As we shall briefly describe, this has led to many predictions which will be the subject of experimental investigation -- see also Ref.~\cite{Saito2007} for a comprehensive review.

A critical development in the application of the QMC model to finite nuclei, where the scalar and vector potentials can vary across the finite size of the nucleon, was carried out in 1996. A purely technical issue, which is nevertheless phenomenologically critical, involved the centre of mass correction to the nucleon mass in-medium. In particular, it was established that this is approximately independent of the applied scalar field. The most important conceptual advance in that work was the derivation of the spin-orbit force, with its explicit dependence on the anomalous magnetic moments and having both isoscalar and isovector components. The latter is now realized to be phenomenologically essential for nuclear structure studies. 

In the last decade the model has been employed to calculate  the equation of state of dense matter, with application to the properties of neutron stars. Not only does the scalar polarizability yield many-body forces between nucleons but it also automatically generates many-body forces between all hadrons, including hyperons, all with no new parameters. As a consequence, the model naturally produces neutron stars with maximum masses around two solar masses when hyperons are included \cite{RikovskaStone:2006ta}. Finally, also within the last decade, the derivation of an energy density functional from the QMC model has allowed this novel theoretical approach to be used in serious nuclear structure calculations \cite{Stone:2016qmi,Stone2017d}.

The outline of this review is the following. Sections~\ref{BaryonInExternal} and \ref{qmc} contain an account of the theoretical development of the QMC model, from hadrons immersed in nuclear matter, to low and high density expansions of the energy density functional. Section~\ref{nuclearmatter} surveys application of the QMC motivated models to nuclear matter at a large range of densities, including compact stars and supernova matter. This is followed by Section~\ref{finite}, devoted to applications of the model to the ground state properties of finite nuclei and hypernuclei. Finally, in Section~\ref{beyond} we discuss ways to test the fundamental idea that hadron structure changes in-medium, with a focus on the consequences of the QMC model for the EMC effect and the Coulomb sum rule. Section~\ref{summary} presents a discussion of the current status of the QMC model and outlines potential future developments.
\begin{figure}[t]
\includegraphics[width=10cm]{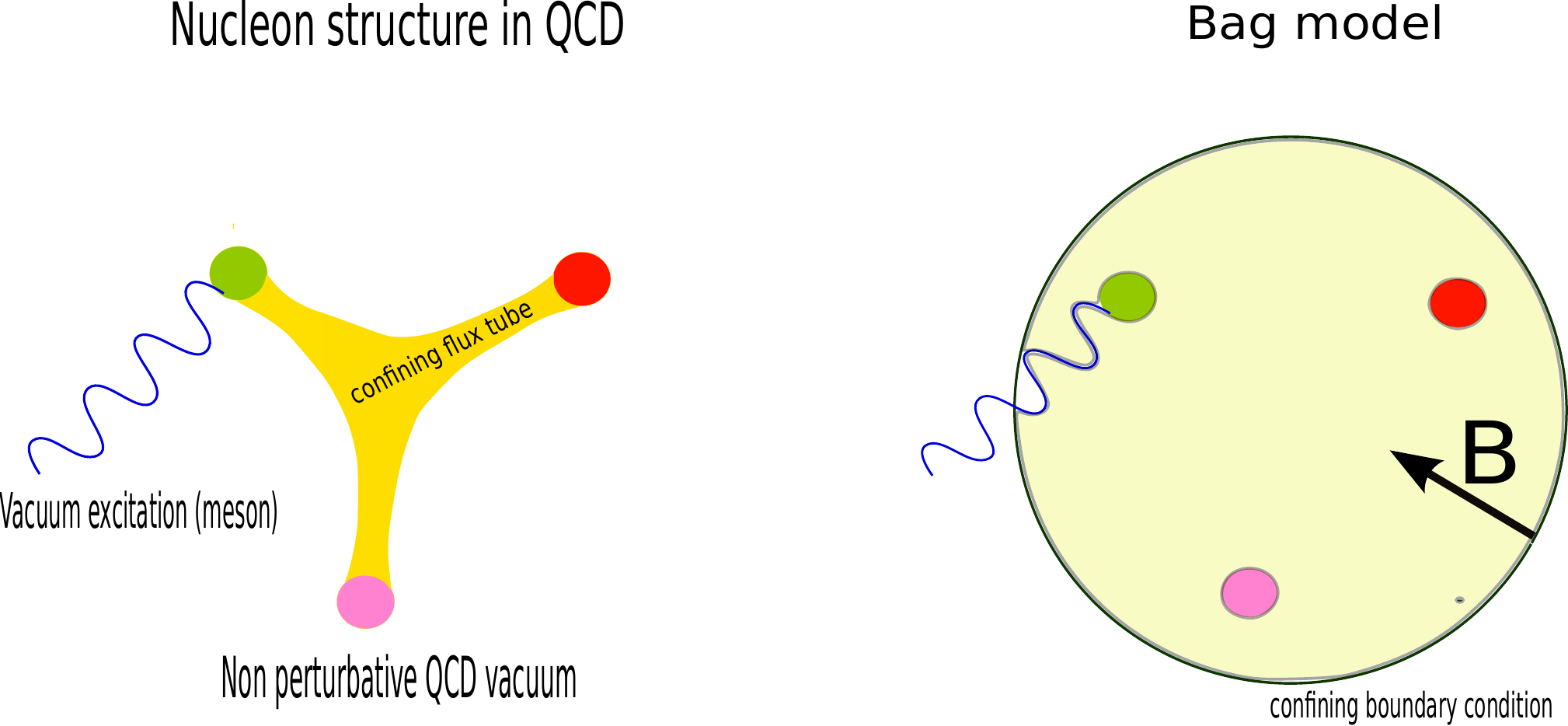}
\centering{}\caption{The QCD picture of the nucleon and the bag model.}
\label{fig:String_Bag_model}
\end{figure}
\section{Baryons in an external field}
\label{BaryonInExternal}

\subsection{A Model for baryon structure}

The salient feature of the QMC model is that the quark structure of
the nucleon plays an essential role in the nuclear dynamics. So we
first introduce and provide some motivation for the bag model, 
which we use to describe the quark
structure of hadrons. Historically the bag \cite{Chodos1974a} was the boundary
between a domain  of perturbative vacuum where the quarks were moving
freely and the non-perturbative vacuum of the QCD. As a consequence the
interaction of the quarks with the external medium was possible only
at the surface of the bag. However the lattice simulations have shown
that this two-phase picture of confinement is misleading. The correct
picture is that the confinement is produced by flux tubes which develop
as the quarks try to escape, as shown 
on the left panel of Fig.\ref{fig:String_Bag_model}.
Inside the tubes the vacuum is approximately perturbative but as they
are rather thin the quark attached at the end obviously feels the
non-perturbative vacuum. In particular, the vacuum excitations, the
mesons, can interact with the quarks. 

Thus, while it is reasonable to describe the quarks as, on average,
being confined in a bag-like region, the interior of the cavity 
can no longer be regarded as having a purely 
perturbative character and the boundary surface is just a device
to account for the fact that the quarks do not escape. The energy
density carried by the flux tubes is diluted over the volume and is
represented by the bag constant $B$. It induces a negative internal
pressure%
\footnote{or positive external pressure%
} which balances the pressure exerted by the confined quarks. Through
this interpretation we recover the historical bag but with the fundamental
difference that the confined quarks can be coupled to the external
meson fields, as shown in the right panel of Fig.\ref{fig:String_Bag_model}. 

In the bag model the quark field, $q(x^{\mu})$, is a solution of the
Dirac equation in free space and satisfies boundary conditions which, for a static
spherical cavity of radius $R_{B}$, takes the form
\begin{eqnarray}
(i\gamma.\partial-m_{q})q & = & 0\,\, r<R_{B},\label{eq:Dirac_eq}\\
(1+i\vec{\gamma}.\hat{x})q & = & 0\,\, r=R_{B}\label{linear_bc},
\end{eqnarray}
where $x^{\mu}=(t,\vec{r})$ and $m_{q}$ is the quark mass. The lowest
positive energy mode with the spin projection $m$ is given as
\begin{equation}
\phi_{m}(\vec{r})=\left(\begin{array}{c}
f(r)\\
i\vec{\sigma}.\hat{r}\, g(r)
\end{array}\right)\frac{\chi_{m}}{\sqrt{4\pi}}={\cal N}\left(\begin{array}{c}
j_{0}\left(xr/R_{B}\right)\\
i\beta_{q}\vec{\sigma}.\hat{r}j_{1}\left(xr/R_{B}\right)
\end{array}\right)\frac{\chi_{m}}{\sqrt{4\pi}}, \label{eq:Lowest_mode}
\end{equation}
where $j_{0},j_{1}$ are the spherical Bessel functions and 
\begin{eqnarray}
\Omega & = & \sqrt{x^{2}+\left(m_{q}R_{B}\right)^{2}},\,\,\beta_{q}=\sqrt{\frac{\Omega-m_{q}R_{B}}{\Omega+m_{q}R_{B}}},\label{eq:Omega0}\\
{\cal N}^{-2} & = & 2R_{B}^{3}j_{0}^{2}(x)\left[\Omega(\Omega-1)+m_{q}R_{B}/2\right]/x^{2}.\label{normalisation}
\end{eqnarray}
The boundary condition is satisfied if $x$ is a solution of
\begin{equation}
j_{0}(x)=\beta_{q}j_{1}(x)
\label{lbc_1}
\end{equation}
and the value of $x$ depends on flavor through the mass $m_{q}$.
In this work we limit our considerations to $u,d,s$ flavors.
The quark field of flavor $f$ in the ground state is then
\[
q_{f}(x^{\mu})=e^{-it\Omega_{f}/R_{B}}\sum_{m}b_{mf}\phi_{m,f}(\vec{r}),
\]
 with $b_{mf}^{\dagger}$ being the creation operator of a quark with spin $m$ and 
flavor $f$.%
\footnote{To avoid confusion the quark flavor will be labeled  $f=u,d,s$ and the
octet baryon flavour $b=p,n,\Lambda,\Sigma^{-},\Sigma^{0},\Sigma^{+},\Xi^{-},\Xi^{0}$.
} The energy of a quark bag with the flavor content $N_{u},N_{d},N_{s}$
is 
\begin{equation}
E=\frac{N_{u}\Omega_{u}+N_{d}\Omega_{d}+N_{s}\Omega_{s}}{R_{B}}+BV.
\label{eq:Ebag_0}
\end{equation}
The volume $V$  of the bag and its radius $R_B$ are determined by the
stability condition 
\begin{equation}
\frac{\partial E}{\partial R_{B}}=0,
\label{eq:stability}
\end{equation}
which implies that the radius is not a free parameter once the bag
constant has been fixed. Note that in practice one often does the reverse, choosing
the radius and fixing $B$ by the stability condition. 

The energy
(\ref{eq:Ebag_0}) cannot yet be identified with a corresponding
hadron. It must be corrected for the zero point motion associated with the
fixed cavity approximation, which takes the form $-Z/R_{B}$, where $Z$
is typically of the order of 3 and is considered to be a free parameter. However,
the expression for the bag energy is still incomplete because it does not depend on the spin
of the particle. In other words, it has $SU(6)$ symmetry, which
is badly violated as  can be seen by comparing the nucleon mass (938MeV)
with the $\Delta_{33}$ resonance energy (1232MeV). In the bag model
this violation is interpreted as a color hyperfine effect, $\Delta_{M}$, 
associated with one gluon exchange \cite{DeGrand1975}. Physically
it corresponds to the interaction of the magnetic moment of one quark
with the color magnetic field created by another quark. Using standard
methods of electromagnetism one finds
\begin{eqnarray}
\Delta E_{M} & = & -3\alpha_{c}\sum_{a,i<j}\lambda_{i}^{a}\lambda_{j}^{a}\vec{\sigma_{i}}.\vec{\sigma_{j}}\frac{\mu_{i}(R_{B})\mu_{j}(R_{B})}{R_{B}^{3}}I_{ij},\label{eq:hyperfine}\\
I_{ij} & = & 1+\frac{2R_{B}^{3}}{\mu_{i}(R_{B})\mu_{j}(R_{B})}\int_{0}^{R_{B}}dr\frac{\mu_{i}(r)\mu_{j}(r)}{r^{4}}
\label{eq:BigInt}
\end{eqnarray}
with
\begin{equation}
\mu_{i}(r)=\frac{r}{6}\frac{4\Omega_{i}+2rm_{i}-3}{2\Omega_{i}\left(\Omega_{i}-1\right)+rm_{i}}.
\label{eq:magnetic-density}
\end{equation}
Here $\lambda^{a}$ are the Gellmann SU(3) matrices and $\alpha_{c}$
is the color coupling constant. We refine this result by taking into
account the fact that the quark wave functions used to get Eq.~(\ref{eq:hyperfine})
do not include the correlation created by the gluon exchange. According to
Ref.~\cite{Barnes1984} the overlap integral in Eq.~(\ref{eq:BigInt}) must
be multiplied by a correction factor $F_{i}F_{j}$, associated with correlations 
generated by higher order gluon exchange. Since the constant
$\alpha_{c}$ is a free parameter,  $F_{u,d}$ can be set to one and we are
left with $F_{s}$ as a free parameter.

To summarize, the mass of a particle takes the form 
\[
M=\frac{N_{u}\Omega_{u}+N_{d}\Omega_{d}+N_{s}\Omega_{s}}{R_{B}}+BV-\frac{Z_{0}}{R_{B}}+\Delta E_{M}
\]
\textcolor{black}{with the free parameters $(B,Z_{0},m_{u},m_{d},m_{s},F_{s},\alpha_{c})$
that can be fixed as follows. First, we set $m_{u}=m_{d}=0$, which is sufficient
for our purpose. (Numerical studies have shown that there is no qualitative change 
if one uses constituent masses, generated by spontaneous chiral symmetry breaking, 
instead of the current masses.) Next we set the free nucleon radius $R_{N}^{free}$
and the nucleon and the $\Delta$ mass equal to their physical values.
Together with the stability equation (\ref{eq:stability}) this determines $B,Z_{0}$ and $\alpha_{c}$.
Finally we choose $m_{s}, F_{s}$ to obtain the best fit to the $(\Lambda,\Sigma,\Xi)$
masses. The results are shown in Table~\ref{FreeParameters} where
the first line corresponds to the fit with $F_{s}=1$. }
\begin{table}
\centering
\caption{\label{FreeParameters}Values of the parameters $(B,Z_{0},m_{s},F_{s},\alpha_{c})$
for $R_{N}^{free}=1.0~\mathrm{fm}$. }
\begin{tabular}{|c|c|c|c|c|c|c|c|}
\hline 
$F_{s}$ & $B(\mathrm{fm}^{-4})$ & $Z_{0}$ & $\alpha_{c}$ & $m_{s}$(MeV) & $M_{\Lambda}$(MeV) & $M_{\Sigma}$(MeV) & $M_{\Xi}$(MeV)\tabularnewline
\hline 
\hline 
1 & 0.284 & 1.771 & 0.560 & $326$ & $1132$ & $1180$ & 1353\tabularnewline
\hline 
0.758  & 0.284 & 1.771 & 0.560 & 289 & 1107 & 1189 & 1325\tabularnewline
\hline 
\hline 
Exp. &  &  &  &  & 1116 & 1195 & 1315\tabularnewline
\hline 
\end{tabular}
\end{table}

\subsubsection{Choosing the bag radius}

A further improvement of the model could be to make it chiral symmetric
by coupling the quarks to pions \emph{\`a la Weinberg}  \cite{Weinberg1969b}\emph{
}as in the Cloudy bag model \cite{Thomas1981,Thomas1984}. However,
for a large bag radius the corrections to the QMC model 
arising from the pion cloud would be rather
moderate, leading simply to a small readjustment of parameters.
Moreover, the pionic corrections to the mass are essentially the same
in the vacuum and in nuclear medium, except for Pauli blocking,
which will be  computed explicitly, see Section~(\ref{Pion-LDA}). 
Therefore we do not attempt such
an improvement which would furthermore seriously complicate the model. 

\begin{figure}[t]
\centering
\includegraphics[height=7cm,width=6cm]{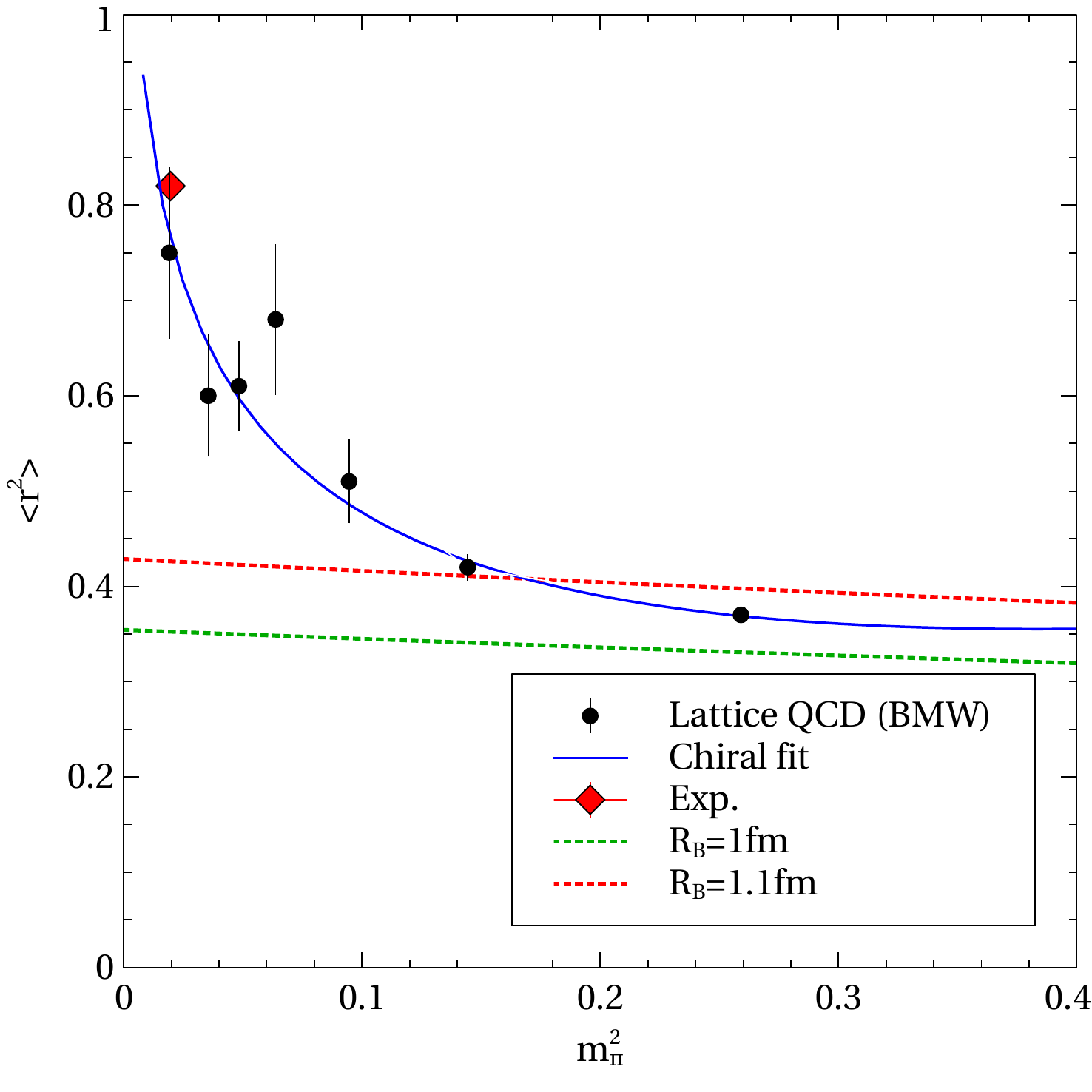}
\caption{Adjustment of the bag radius to lattice QCD calculations.}
\label{fig:MatchLattice}
\end{figure}

However, to set the bag radius, we cannot compare directly the squared
charge radius of the bag to the experimental value, since it
is measured in a world with approximate chiral symmetry that the model
breaks. To get around the problem we use lattice QCD calculations
to extrapolate the physical value of the charge radius to the value it would have, in a non-chiral
symmetric world, that is with a large pion mass. This value can be then compared
to the prediction of a bag with large quark masses, for which
chiral effects are certainly negligible. This procedure is illustrated for the isovector
squared radius in Fig.~\ref{fig:MatchLattice}. 
The lattice calculation \cite{Metivet:2015alr},
which used the  gauge ensembles of the  Budapest-Marseille-Wuppertal collaboration, covers a large range of pion masses,
down to the physical one. The continuous blue curve is the chiral
fit proposed in Ref.~\cite{Hall2013}. The good agreement with the
experimental value $0.82~\mathrm{fm}^{2}$ at the physical pion mass tells us
that the lattice results for large pion masses should be reliable.
The red and green lines are the bag model prediction as a function
of the quark mass. The lattice results thus suggest that the bag radius
should be around 1fm. This is the value that we shall adopt.

\subsection{Bag in an external field}

In the QMC model we assume that the interactions arise via the coupling
of  quarks to meson fields. The dominant exchanges are the scalar
($\sigma)$, which is the origin of the intermediate range attraction and the
vector ($\omega)$, which provides the short range repulsion. Both
are isospin independent. The vector isovector ($\rho)$ exchange  will be introduced
later in Section~\ref{rho}. The time dependence of the meson field can be neglected as
the nuclear excitation energies are much lower than the meson mass.
Moreover, we shall also neglect the space component of the vector field
because there is no available vector in nuclear matter to set the
direction%
\footnote{strictly speaking this is only true for the average of the field. %
}. Thus we write $\omega^{\mu}=(\omega,\vec{\omega}\sim0).$ 

Let us consider a baryon bag at rest located at the origin and interacting
with given $(\sigma,\omega)$ classical (that is C-number) fields.
Then its energy is $E_{Bag}=\langle H_{Bag}\rangle$ with $\langle...\rangle$
denoting the expectation value in the quark Fock space  and 
\begin{equation}
H_{Bag}=\sum_{f}\int^{R_{B}}d\vec{r}\left[-i\bar{q}_{f}\gamma_{0}\vec{\gamma}.\vec{\nabla}q_{f}+(m_{f}-g_{\sigma}^{qf}\sigma)\bar{q}_{f}q_{f}+g_{\omega}^{qf}\omega q_{f}^{\dagger}q_{f}\right]\label{eq:H_bag}
\end{equation}
corresponding to the Dirac equation in the presence of the meson fields%
\footnote{Note that the boundary condition is not changed by the coupling to
the mesons%
}:
\begin{eqnarray}
(i\gamma.\partial-m_{f}+g_{\sigma}^{qf}\sigma-g_{\omega}^{qf}\gamma_{0}\omega)q_{f} & = & 0\,\, r<R_{B},\label{eq:Dirac_eq-1}\\
(1+i\vec{\gamma}.\hat{x})q_{f} & = & 0\,\, r=R_{B}\label{linear_bc-1},
\end{eqnarray}
where $m_f$ and $q_f$ are the mass and wave function of a quark with flavor $f$  and $g_{\sigma}^{qf}, g_{\omega}^{qf}$ are the quark-meson coupling constants.

\subsubsection{Constant field}

\begin{figure}
\centering{}\includegraphics[height=7cm,width=6cm]{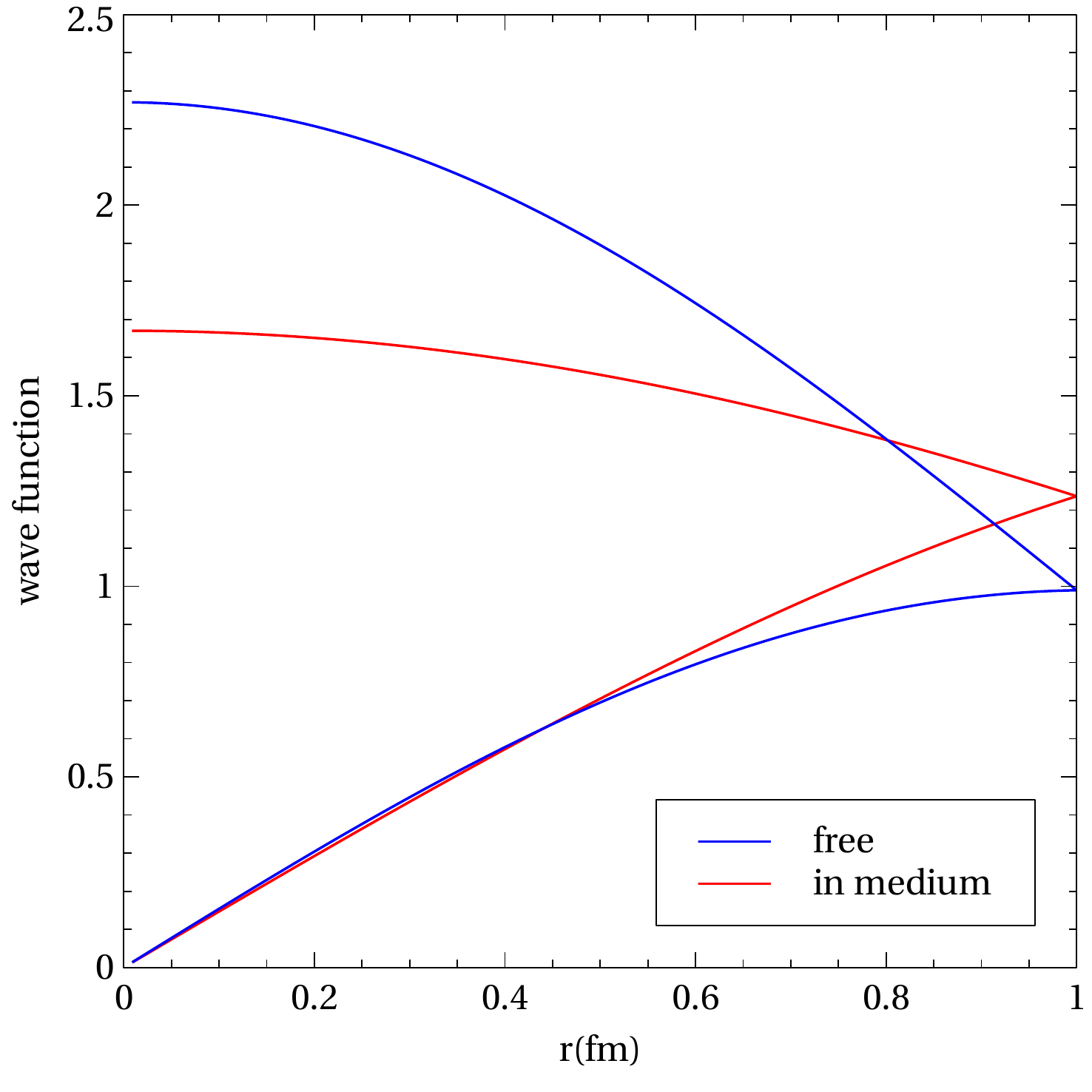}
\caption{The upper and lower component of the quark wave function in the free case and in the medium, for a typical value of the $\sigma$ field.}
\label{fig:QuarkWF}
\end{figure}

Let us assume that it makes sense to consider that the meson fields
are constant over the volume of the bag. Then
\[
\sum_{f}\int^{R_{B}}d\vec{r}g_{\omega}^{qf}\omega q_{f}^{\dagger}q_{f}=\omega\sum_{f}N_{f}g_{\omega}^{qf}
\]
and we can define a baryon-omega coupling
\[
g_{\omega}^{b}=\sum_{f}N_{f}g_{\omega}^{qf}.
\]
 The energy of a baryon bag in the external field is 
\begin{eqnarray*}
M_{b} & = & M_{b}(\sigma)+g_{\omega}^{b}\omega ,\\
M_{b}(\sigma) & = & \sum_{f}\frac{N_{f}\Omega_{f}(m_{f}-g_{\sigma}^{qf}\sigma)}{R_{B}}+BV-\frac{Z_{0}}{R_{B}}+\Delta E_{M}( m_{f}-g_{\sigma}^{qf}\sigma).
\end{eqnarray*}
While the $\omega$ exchange yields  only an additive contribution to the mass, the $\sigma$ exchange changes
the internal structure of the bag by making the quark ``more relativistic''.
We illustrate these features in Fig.\ref{fig:QuarkWF} where we show how $g(r)$, the
lower component of the quark wave function, increases as the $\sigma$ field grows. This change
of the internal structure of the bag has an important consequence for nuclear physics because
it implies that the effective $\sigma$-baryon coupling, proportional to
the scalar integral:
\[
I_{s}=\int^{R_{B}}d\vec{r}\bar{q}q\sim\int^{R_{B}}d\vec{r}\left(f^{2}-g^{2}\right)
\]
decreases with increasing applied $\sigma$ field. {\em This is basis of the quark mechanism
for the saturation of nuclear forces.}

The expression for $M_{b}(\sigma)$ is somewhat inconvenient for use
in actual calculations. A series expansion in powers of $\sigma$
is more useful. We have chosen to write it in the form 
\begin{eqnarray*}
M_{b}(\sigma) & = & M_{b}-g_{\sigma}^{b}\sigma+\frac{d_{b}}{2}\left(g_{\sigma}^{b}\sigma\right)^{2}+...\\
 & = & M_{b}-w_{b}g_{\sigma}\sigma+\frac{d}{2}\tilde{w}_{b}\left(g_{\sigma}\sigma\right)^{2},
\end{eqnarray*}
where $g_{\sigma}=g_{\sigma}^{N}$ is the (free) nucleon-meson coupling
constant. In the following we prefer the second form because the weights $w,\tilde{w}$
give directly the relative strength of the strange particle couplings.
Note that the coupling defined as the slope of $M_{b}(\sigma)$ is
actually equal to that defined in Eq.~(\ref{eq:H_bag}) by the scalar
integral $I_{s}$ and the $\sigma-{\rm quark}$ couplings%
\footnote{provided one neglects $\Delta E_{M}$%
}. The coefficient $d$ leads to the reduction of the effective $\sigma$-N coupling
as the applied $\sigma$ field (and of course the baryon density) increases. 
By analogy with electromagnetism
we call it the \textit{scalar polarizability}. Since we set $m_{u}=m_{d}=0$, 
there is no distinction between proton and neutron so $d$ refers
to the nucleon and by definition its weights are equal to 1. For the
same reason there is no distinction between the isospin partners of
the $\Sigma$ and $\Xi.$ The fit of the coefficients $\{d,w,\tilde{w}\}$
has been done using the parameters given in Table~\ref{FreeParameters}, 
with the result shown in Table~\ref{weigths}.
\begin{table}
\centering
\caption{\label{weigths} The weights and polarizability from the fit.}
\begin{tabular}{|c|c|c|c|c|}
\hline 
   & $N$ & $\Lambda$ & $\Sigma$ & $\Xi$\tabularnewline
\hline 
\hline 
$w$ & 1 & 0.7114 & 0.5847 & 0.3400\tabularnewline
\hline 
$\tilde{w}$ & 1 & 0.68956 & 0.6629 & 0.3688\tabularnewline
\hline 
$d$ & 0.1797 &  &  & \tabularnewline
\hline 
\end{tabular}
\end{table}

It turns out that this quadratic expansion is sufficiently accurate
for our purposes, even for values of the field reached in massive
neutron stars.

\subsubsection{ Slowly varying field and a moving bag}

Here we derive the results which are pertinent for finite nuclei. 
In particular we  explain in some detail how the spin-orbit force appears in the QMC model.

The case of a bag moving in a constant field is trivial since the
change with respect to the bag at rest may be taken into account by a
simple boost, with the result
\[
E_{b}=\sqrt{P^{2}+M_{b}^{2}(\sigma)}+g_{\omega}^{b}\omega.
\]
We now consider the case of a bag moving in meson fields which vary slowly 
as a function of position.
This is relevant only for finite nuclei, since for applications of
the model to neutron stars the approximation of uniform matter is sufficient. 

We summarise here our previous work~\cite{Guichon1996}, where the reasoning
was presented in full. Following the Born-Oppenheimer approximation
it was argued that the motion of the bag and the variation of the
fields were slow enough that the motion of the quarks, which is relativistic, 
could adjust instantaneously
to the actual value of the fields. The justification used the fact
that in a nucleus the meson fields essentially follow the nuclear
density. If one denotes by $\vec{R}$ the position of the bag, then in
the first approximation one can neglect variation of the field over
the volume and the statement that the quarks adjust their motion 
to the actual value of the fields amounts
to stating that they stay in the lowest orbit corresponding to the
fields $\sigma(\vec{R}), \omega(\vec{R)}$. This leads to the obvious
result
\[
E_{b}(\vec{R})=\sqrt{P^{2}+M_{b}^{2}(\sigma(\vec{R)})}+g_{\omega}^{b}\omega(\vec{R}).
\]

Next let us consider the first order correction associated with the variation
of the field over the volume of the bag. As explained in
Ref.~\cite{Guichon1996}, in the instantaneous rest frame (IRF) of
the bag with velocity $\vec{v}$ , the $\omega$ field has a space
component $\sim\omega\vec{v}$ . This component induces a 
magnetic field $\vec{B}\sim\vec{\nabla}\times(\omega\vec{v})$, 
which is non-zero at the surface of the nucleus. This field leads to the interaction 
$-\vec{\mu}.\vec{B}$. The quantity, $\mu$, appearing
here, is exactly the same as that for the isoscalar magnetic moment, 
evaluated in the mean-field $\sigma$, in the MIT bag model.   
Thus the isoscalar magnetic moment of the nucleon 
naturally determines the strength of the isoscalar 
part of the spin-orbit interaction. Assuming this is a small
correction, it was computed in Ref.~\cite{Guichon1996} as a perturbation
with the result: 
\begin{equation}
V_{Magn}^{b}=\frac{1}{MM_{b}(\sigma)}\frac{\mu_{IS}(\sigma)}{\mu_{N}}\vec{S}.\vec{\nabla}\left(g_{\omega}^{b}\omega\right)\times\vec{P},\label{eq:H_SO_magnetic}
\end{equation}
with $M, \vec{S}, \vec{P}$ being the free nucleon mass, 
the spin and the momentum of the baryon, respectively.
Of course, in-medium the isoscalar magnetic moment $\mu_{IS}(\sigma)$ 
varies with the $\sigma$ field as the integral
\[
\int^{R_{B}}r^{3}drf(r)g(r)
\]
with $f,g$ computed in the $\sigma$ field. The value for the nucleon
at $\sigma=0$ is taken from experiment $\mu_{IS}(0)=0.88\mu_{N}.$

However, the expression (\ref{eq:H_SO_magnetic}) is only a part 
of the full spin-orbit interaction. Let us suppose
that the bag moves along some trajectory under the influence of a
force which does not couple to the spin which therefore remains constant.
This means that the spin components of the baryon in the IRF$(t)$ satisfy $S^{i}(t+dt)=S^{i}(t$).
But what we need are the components of the spin at $t+dt $ in the IRF$(t+dt)$
which must be different from IRF$(t)$ if the bag accelerates along the trajectory.
If we denote by $\Lambda(\vec{v})$ the Lorentz transformation from the nuclear
rest frame to the IRF with velocity $\vec{v}$, then the spin components
in the IRF$(t+dt)$ are 
\[
\Lambda(\vec{v}+d\vec{v})\Lambda^{-1}(\vec{v})S^{i}(t+dt)=\Lambda(\vec{v}+d\vec{v})\Lambda^{-1}(\vec{v})S^{i}(t),
\]
where $\vec{v}+d\vec{v}$ is the velocity of the IRF at $t+dt.$
The product $\Lambda(\vec{v}+d\vec{v})\Lambda^{-1}(\vec{v})$ is a
boost times a rotation called Thomas precession. In order to 
describe this effect by the Hamilton equation of motion, a new spin orbit
interaction must be added to the QMC Hamiltonian. The complete derivation can be found
in Ref.~\cite{Jackson1998}. Here we propose a simple trick to get this
precession term. 

Since the Thomas precession is a kinematic effect, 
independent of the baryon structure, it can be 
derived from the point-like Dirac equation in a scalar field. Since this field
does not couple to the spin, the result will effectively be the Thomas
precession. To exhibit the corresponding spin-orbit interaction one
performs the Foldy-Wouthuysen transformation (FW) \cite{Foldy1950} of the
interaction so that it incorporates the leading relativistic effects.
It can then be used in a non-relativistic approximation. If we write
the Dirac equation in the form: 
\begin{eqnarray*}
i\frac{\partial}{\partial t}\psi & = & \left[\gamma_{0}\vec{\gamma}.\vec{p}+\gamma_{0}(m+s(r)\right]\psi\\
 & = & \left[\gamma_{5}\vec{\sigma}.\vec{p}+\gamma_{0}(m+s(r)\right]\psi,
\end{eqnarray*}
then the FW transformation of the interaction is \cite{Bjorken1964}
\[
-\frac{1}{8m^{2}}\left[\gamma_{5}\vec{\sigma}.\vec{p},\left[\gamma_{5}\vec{\sigma}.\vec{p},\gamma_{0}s(r)\right]\right]=-\gamma_{0}\frac{1}{4m^{2}}\vec{\sigma}.\vec{\nabla}s\times\vec{p}+{\rm spin\, independent\, part.}
\]
In a non-relativistic approximation $\gamma_{0}$ can be set to 1
and since the Pauli matrix represents twice the spin we get the general
result that the precession spin-orbit is 
\[
\frac{1}{2m^{2}}\vec{S}.\vec{F}\times\vec{p},
\] 
where $\vec{F}=-\vec{\nabla}s$ is the force driving the motion.

If we were to do the same with a vector interaction, $w(r),$ we would
get: 
\[
-\frac{1}{8m^{2}}\left[\gamma_{5}\vec{\sigma}.\vec{p},\left[\gamma_{5}\vec{\sigma}.\vec{p},w(r)\right]\right]=+\frac{1}{4m^{2}}\vec{\sigma}.\vec{\nabla}w\times\vec{p}+{\rm spin\, independent\, part},
\]
which is the same as the scalar one but with opposite sign. This does
not mean that the precession effect changes sign when going from the
scalar to the vector interaction. This result is in fact the sum of
the precession and of the interaction of the magnetic moment with
the rest frame magnetic field. What is misleading is that, for a 
point-like particle, the magnetic term is just about twice the precession term,
as one can check from Eq.~(\ref{eq:H_SO_magnetic}), 
where one sets $\mu_{IS}(\sigma)=\mu_{N}$.
Thus the precession term is actually the same for the scalar and the
vector interaction, as it must be since this is a geometric effect
which does not depend on the nature of the applied force. 
In the QMC
model this force is $-\vec{\nabla}\left(M_{b}(\sigma)+g_{\omega}^{b}\omega\right)$, 
so we can write: 
\begin{equation}
V_{prec}^{b}=-\frac{1}{2M_{b}^{2}(\sigma)}\vec{S}.\vec{\nabla}\left(M_{b}(\sigma)+g_{\omega}^{b}\omega\right)\times\vec{P} \, .
\label{eq:H_S)_precession}
\end{equation}

\subsubsection{Interaction with the $\rho$ meson field }
\label{rho}

The interaction with an isovector field $\vec{B}$%
\footnote{we cannot label it by $\rho$ since that is reserved for the nuclear baryonic 
density.%
} $(B^{\alpha},\alpha=1,2,3)$ is introduced by adding the term
\[
g_{\rho}q_{f}^{\dagger}\frac{\vec{B}.\vec{\tau}}{2}q_{f}
\]
 to the quark bag energy (\ref{eq:H_bag}), where the flavor Pauli
matrices $\vec{\tau}$ carry all the flavor dependence of the coupling.
The leading term for the energy of a baryon $b$ in the applied fields
then becomes 
\[
E_{b}(\vec{R})=\sqrt{P^{2}+M_{b}^{2}(\sigma(\vec{R)})}+g_{\omega}^{b}\omega(\vec{R})+g_{\rho}\vec{I}.\vec{B}(\vec{R}),
\]
 where $\vec{I}$ is the isospin of the particle (see the Appendix). The
spin-orbit interaction becomes
\begin{equation}
V_{Magn}^{b}=\frac{1}{MM_{b}(\sigma)}\left(\frac{\mu_{IS}(\sigma)}{\mu_{N}}\vec{S}.\vec{\nabla}g_{\omega}^{b}\omega+\frac{\mu_{IV}(\sigma)}{\mu_{N}}\vec{S}.\vec{\nabla}g_{\rho}\vec{B}.\vec{I}\right)\times\vec{P} \, , 
\label{eq:Vso-magn}
\end{equation}
with $\mu_{IV}(0)=4.7\mu_{N}$ and 
\begin{equation}
V_{prec}^{b}=-\frac{1}{2M_{b}^{2}(\sigma)}\vec{S}.\vec{\nabla}\left(M_{b}(\sigma)+g_{\omega}^{b}\omega+g_{\rho}\vec{B}.\vec{I}\right)\times\vec{P} \, .
\label{eq:Vso-precess}
\end{equation}

\subsection{The meson Hamiltonian }
\label{meson}

For completeness we discuss now the Hamiltonian of the mesons. We
postpone the case of the pion field, which will be treated later as
perturbation to Section~(\ref{Pion-LDA}). For the scalar field the Hamiltonian has the form
\[
E_{\sigma}=\int d\vec{r}\left[\frac{1}{2}\left(\vec{\nabla}\sigma\right)^{2}+V(\sigma)\right],
\]
where the potential energy, $V(\sigma)=m_{\sigma}^{2}\sigma^{2}/2+\cdots$, 
is generally limited to the quadratic term. This has been the case in our
previous work~\cite{Guichon2004,Guichon:2006er,RikovskaStone:2006ta,Stone:2016qmi}.
However, the scalar polarizability $d$ can be seen as the zero energy
limit of the $\sigma-N$ scattering amplitude and we know from general
principles that it must have a pole in the $t$ channel variable.
This pole would appear as a divergence of the sum over $s$-channel
intermediate states but when one computes $d$ using the bag equations
this divergence is of course not present since it is in the unphysical
region. In a dispersive langage one would call this a subtraction.

\begin{figure}
\centering{}\includegraphics[clip,width=12cm]{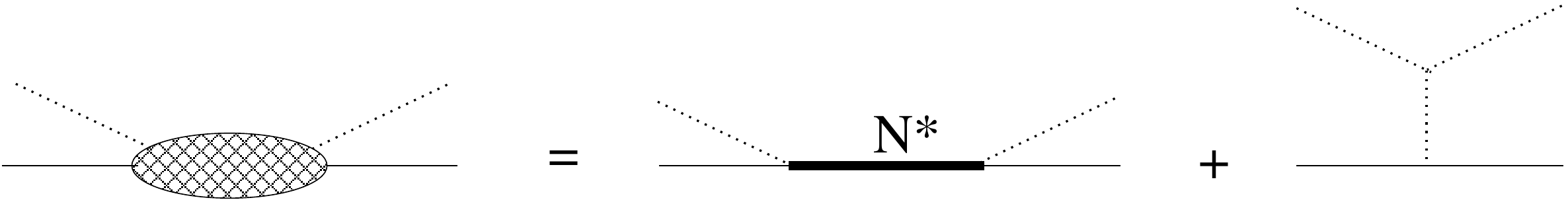}
\caption{$s$-channel and $t$-channel contribution to the scalar polarizability $d$.}
\label{fig:scalar-polar}
\end{figure}

\begin{figure}
\centering{}\includegraphics[clip,width=12cm]{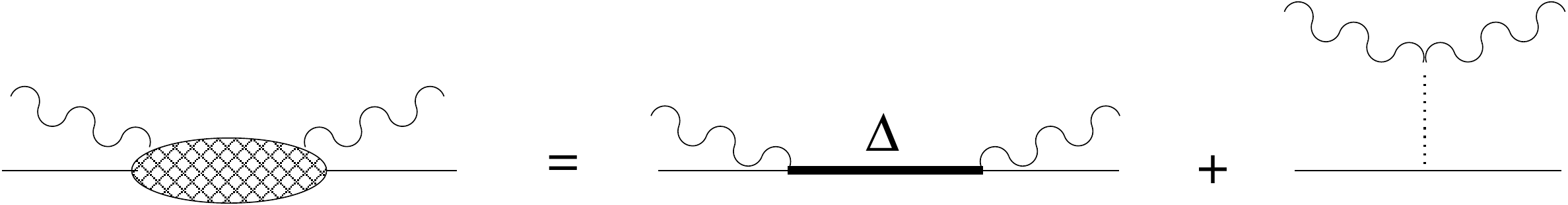}
\caption{$s$-channel and $t$-channel contribution to the magnetic polarizability.}
\label{fig:magnetic-polar}
\end{figure}

To restore this piece of the amplitude we must supplement the model
with a $\sigma^{3}$ interaction, which will create the $t$-channel
pole as illustrated in Fig.\ref{fig:scalar-polar}.
Note that the same mechanism is at work for the magnetic polarizability, 
where the $s$-channel contribution (dominated by the $\Delta$ or
$P_{33}$ resonance) is partly cancelled by the $t$-channel exchange
of the $\sigma$ meson (see Fig.~\ref{fig:magnetic-polar}).
In this case one must introduce a $\sigma\gamma\gamma$ interaction
to generate the pole.
Thus we can write for the scalar field
\begin{eqnarray*}
V(\sigma) & = & \frac{m_{\sigma}^{2}\sigma^{2}}{2}+\frac{\lambda_{3}}{3!}\left(g_{\sigma}\sigma\right)^{3}+\frac{\lambda_{4}}{4!}\left(g_{\sigma}\sigma\right)^{4}.
\end{eqnarray*}
The quartic term has been added to guarantee the existence of a ground
state. The constant $\lambda_{4}$ may be arbitrarily small but must be positive.
For vector fields there is no such motivation and we use the simple
form, appropriate for the space component of a 4-vector field:
\[
E_{\omega,\rho}=-\frac{1}{2}\int d\vec{r}\left[\left(\vec{\nabla}\omega\right)^{2}+m_{\omega}^{2}\omega^{2}+\vec{\nabla}B_{\alpha}.\vec{\nabla}B_{\alpha}+m_{\rho}^{2}B_{\alpha}B_{\alpha}\right].
\]

\section{QMC model}
\label{qmc}

\subsection{The full model}

We assume that the nuclear system can be treated as a collection of
non-overlapping bags, so that its total energy is simply
\begin{equation}
E_{QMC}=\sum_{i=1,...}\sqrt{P_{i}^{2}+M_{i}^{2}(\sigma(\vec{R}_{i})}+g_{\omega}^{i}\omega(\vec{R}_{i})+g_{\rho}\vec{I}_{i}.\vec{B}(\vec{R}_{i})+E_{\sigma}+E_{\omega,\rho},\label{eq:E_QMC}
\end{equation}
where $\vec{R}_{i},\vec{P}_{i}$ are the position and momentum of
baryon $i$. The spin-orbit interaction, which has been derived
as a perturbation, and the single pion exchange are not included in 
the following derivation and will be added at the end. To simplify
the presentation, the label $i$, which numbers the baryons present in
the system, implicitly specifies the flavor $b$ which would normally
be written as $b(i).$

Since the meson mean-fields are time independent one can eliminate them through
the equations of motion:
\[
\frac{\delta E_{QMC}}{\delta\sigma(\vec{r})}=\frac{\delta E_{QMC}}{\delta\omega(\vec{r})}=\frac{\delta E_{QMC}}{\delta B_{\alpha}(\vec{r})}=0 \, .
\]
That is
\begin{eqnarray}
-\nabla_{r}^{2}\sigma+\frac{dV}{d\sigma(\vec{r})} & = & -\sum_{i}\delta(\vec{r}-\vec{R}_{i})\frac{\partial}{\partial\sigma}\sqrt{P_{i}^{2}+M_{i}^{2}(\sigma(\vec{R}_{i})},\label{eq:sigma-Eq}\\
\nabla_{r}^{2}\omega-m_{\omega}^{2}\omega(\vec{r}) & = & -g_{\omega}\sum_{i}\delta(\vec{r}-\vec{R}_{i})w_{\omega}^{i}=-g_{\omega}D_{\omega}(\vec{r}),\label{eq:omega-Eq}\\
\nabla_{r}^{2}B_{\alpha}-m_{\rho}^{2}B_{\alpha}(\vec{r}) & = & -g_{\rho}\sum_{i}\delta(\vec{r}-\vec{R}_{i})I_{\alpha}^{i}=-g_{\rho}D^{\alpha}(\vec{r}) \, , 
\label{eq:rho-Eq}
\end{eqnarray}
where we have set the weights $w_{\omega}^{i}=g_{\omega}^{i}/g_{\omega}$ with
$g_{\omega}=g_\omega^N$ being the coupling constant of the nucleon. 

Consistent with the Zweig rule, within the QMC model the $\sigma, \omega$ and $\rho$ mesons  
(which contain no strange quarks) are taken to couple {\em only} to the $u$ and $d$ quarks. Thus, 
for baryons we find 
\[
w_{\omega}^{i}=\frac{g_{\omega}^{i}}{g_{\omega}}=1+\frac{s(i)}{3},
\]
where $s(i)$ is the strangeness of particle $i$. For the $\rho$ meson the coupling is determined 
by the isospin operator, $I$, whether acting on a nucleon or a hyperon (see the Appendix) 
and so, for example, its coupling to the $\Lambda$ hyperon vanishes.

Once the equations (\ref{eq:sigma-Eq}) - (\ref{eq:rho-Eq}) for the meson fields are solved, one just
needs to substitute them in $E_{QMC}$ (\ref{eq:E_QMC}) to get the canonical \emph{classical}
Hamiltonian for the interacting baryons. The system is then quantized 
by the replacement 
\[
\vec{P}_{i}\to-i\vec{\nabla}_{i}.
\]
As the equations for the $\omega,\rho$ fields are linear, this procedure leads simply to the standard 2-body Yukawa potential: 
\begin{eqnarray}
H_{\omega} & = & \frac{1}{2}g_{\omega}^{2}\int d\vec{r}d\vec{r}'\, D_{\omega}(\vec{r})\langle\vec{r}\left|\left(-\nabla_{r}^{2}+m_{\omega}^{2}\right)^{-1}\right|\vec{r}'\rangle D_{\omega}(\vec{r}'),\label{eq:H_omega}\\
H_{\rho} & = & \frac{1}{2}g_{\rho}^{2}\int d\vec{r}d\vec{r}'\, D^{\alpha}(\vec{r})\langle\vec{r}\left|\left(-\nabla_{r}^{2}+m_{\rho}^{2}\right)^{-1}\right|\vec{r}'\rangle D^{\alpha}(\vec{r}').\label{eq:H_rho}
\end{eqnarray}

The heavy $\sigma, \omega$ and $\rho$ mesons account for the exchange of correlated pions, but
the single pion exchange must be added separately.
Since this exchange does not have a mean field value it comes into play only
through the fluctuations and is thus a rather small contribution.
At the quark level the coupling is motivated by chiral symmetry which
imposes a derivative coupling. At the baryon level this leads to the
interaction
\[
\frac{3g_{A}}{5f_{\pi}}\int d\vec{r}\vec{j}_{\pi}^{\alpha}(\vec{r}).\vec{\nabla}\phi^{\alpha}(\vec{r}) \, ,
\]
with the source term
\[
\vec{j}_{\pi}^{\alpha}(\vec{r})=\sum_{i}\delta\left(\vec{r}-\vec{R}_{i}\right)\vec{G}_{T}^{\alpha}(i)
\]
and $\vec{G}_{T}^{\alpha}$ being the Gamow-Teller operator acting on the
baryon $i$ (see the Appendix). We use $g_{A}=1.26,\, f_{\pi}=0.9$ MeV.
Eliminating the static pion field leads to the pionic interaction 
\begin{equation}
H_{\pi}=\frac{1}{2}\left(\frac{3g_{A}}{5f_{\pi}}\right)^{2}\int d\vec{r}d\vec{r}'\,\vec{j}_{\pi}^{\alpha}(\vec{r}).\vec{\nabla_{r}}\,\vec{j}_{\pi}^{\alpha}(\vec{r}').\vec{\nabla}_{r'}\langle\vec{r}|\left(-\nabla_{r}^{2}+m_{\pi}^{2}\right)^{-1}|\vec{r}'\rangle.\label{eq:H_pion}
\end{equation}

By contrast, the $\sigma$ meson exchange contribution to the Hamiltonian is 
highly non-trivial, because of the non-linear dependence 
of the RHS of Eq.~\ref{eq:sigma-Eq} on the $\sigma$
field itself. This feature creates N-body interactions which make the Hamiltonian
highly impractical as it stands, not to speak of the problem of defining the square
root operator.

\subsection{Expansion about the mean $\sigma$ field}

In order to obtain a form of the Hamiltonian which can actually be used, 
we assume that it makes sense to write
for the field operator $\sigma$
\[
\sigma=\langle\sigma\rangle+\delta\sigma,
\]
where the \emph{C-number} $\langle\sigma\rangle\equiv\bar{\sigma}$
denotes the ground state expectation value, that is
\begin{equation}
\langle\sigma(\vec{r})\rangle=\int d\vec{R}_{1}...d\vec{R}_{A}\Phi^{*}(\vec{R}_{1}...\vec{R}_{A})\sigma(\vec{r},\vec{R}_{i},\vec{P}_{i})\Phi(\vec{R}_{1}...\vec{R}_{A})\label{eq:Average field}
\end{equation}
and to consider the fluctuation $\delta\sigma$ as a small quantity. 

More precisely, if we define
\begin{equation}
K=\sum_{i}\delta(\vec{r}-\vec{R}_{i})\sqrt{P_{i}^{2}+M_{i}(\sigma)^{2}}\label{eq:Kinetic}
\end{equation}
we see that the meson field equation has the form
\[
-\nabla^{2}\left(\bar{\sigma}+\delta\sigma\right)+\frac{dV}{d\sigma}\left(\bar{\sigma}+\delta\sigma\right)=-\frac{\partial K}{\partial\sigma}=-\frac{\partial K}{\partial\sigma}(\bar{\sigma})-\delta\sigma\frac{\partial^{2}K}{\partial\sigma^{2}}(\bar{\sigma})-\cdots,
\]
where we have used the loose notation
\[
\frac{\partial K}{\partial\sigma}(\bar{\sigma})\equiv\left.\frac{\partial K}{\partial\sigma}\right|_{\sigma=\bar{\sigma}}.
\]
We also expand $\frac{\partial K}{\partial\sigma}(\bar{\sigma})$ 
and $\frac{\partial^{2}K}{\partial\sigma^{2}}(\bar{\sigma})$
about their expectation values
\begin{eqnarray*}
\frac{\partial K}{\partial\sigma}(\bar{\sigma}) & = & \langle\frac{\partial K}{\partial\sigma}(\bar{\sigma})\rangle+\delta\left[\frac{\partial K}{\partial\sigma}(\bar{\sigma})\right],\\
\frac{\partial^{2}K}{\partial\sigma^{2}}(\bar{\sigma}) & = & \langle\frac{\partial^{2}K}{\partial\sigma^{2}}(\bar{\sigma})\rangle+\delta\left[\frac{\partial^{2}K}{\partial\sigma^{2}}(\bar{\sigma})\right]
\end{eqnarray*}
and consider that 
\[
\delta\sigma,\,\delta\left[\frac{\partial K}{\partial\sigma}(\bar{\sigma})\right],\,\delta\left[\frac{\partial^{2}K}{\partial\sigma^{2}}(\bar{\sigma})\right]
\]  
are small quantities. Then the meson field equation can be solved
order by order, which gives:
\begin{eqnarray}
-\nabla^{2}\bar{\sigma}+\frac{dV}{d\sigma}\left(\bar{\sigma}\right) & = & -\langle\frac{\partial K}{\partial\sigma}(\bar{\sigma})\rangle,\label{eq:order 0}\\
-\nabla^{2}\delta\sigma+\frac{d^{2}V}{d\sigma}\left(\bar{\sigma}\right)\delta\sigma & = & -\delta\left[\frac{\partial K}{\partial\sigma}(\bar{\sigma})\right]-\delta\sigma\langle\frac{\partial^{2}K}{\partial\sigma^{2}}(\bar{\sigma})\rangle\nonumber \\
 & = & -\frac{\partial K}{\partial\sigma}(\bar{\sigma})+\langle\frac{\partial K}{\partial\sigma}(\bar{\sigma})\rangle-\delta\sigma\langle\frac{\partial^{2}K}{\partial\sigma^{2}}(\bar{\sigma})\rangle,
\label{eq:order 1}\\
... & = & ...\nonumber 
\end{eqnarray}
As we limit the expansion of the Hamiltonian to order $(\delta\sigma)^{2},$
it is sufficient to solve the field equations at order $\delta\sigma$,
which corresponds to Eqs.~(\ref{eq:order 0}, \ref{eq:order 1}). Starting
from Eq.~(\ref{eq:E_QMC}), after some integration by parts and neglecting
terms of order higher than $(\delta\sigma)^{2},$ we get the $\sigma$
part of the Hamiltonian 
\begin{eqnarray}
H_{\sigma}&=&   \int d\vec{r}\,K(\bar{\sigma})+V(\bar{\sigma})-\frac{1}{2}\bar{\sigma}\left(\langle\frac{\partial K}{\partial\sigma}(\bar{\sigma})\rangle+\frac{dV}{d\sigma}\right)\nonumber\\
          &+&\frac{1}{2}\delta\sigma\left(\frac{\partial K}{\partial\sigma}(\bar{\sigma})-\langle\frac{\partial K}{\partial\sigma}(\bar{\sigma})\rangle\right).\label{eq:H_sigma}
\end{eqnarray}
Note that $\bar{\sigma}$ is a C-number so $V(\bar{\sigma})$ is the
same as $\langle V(\bar{\sigma})\rangle$.

In practice, in order to construct the Hartree-Fock equations we only need the expectation value
in the ground state that is
\begin{equation}
\langle H_{\sigma}\rangle=\int d\vec{r}\,\left[\langle K(\bar{\sigma})+V(\bar{\sigma})\rangle-\frac{1}{2}\bar{\sigma}\left(\langle\frac{\partial K}{\partial\sigma}(\bar{\sigma})\rangle+\frac{dV}{d\sigma}\right)+\frac{1}{2}\langle\delta\sigma\frac{\partial K}{\partial\sigma}(\bar{\sigma})\rangle\right],\label{eq:<H_sigma>}
\end{equation}
where we have used $\langle\delta\sigma\rangle=0.$

To complete the system we need a prescription to write the quantum
form of $K$ and its derivatives. The essential point is that, thanks
to the expansion, we only need $K(\bar{\sigma})$ which is a \emph{one
body operator} because $\bar{\sigma}$ is a C-number. So we can write
\[
K(\bar{\sigma})=\sum_{i}\delta(\vec{r}-\vec{R}_{i})\sqrt{P_{i}^{2}+M_{i}[\bar{\sigma}(\vec{r})]^{2}}=\sum_{\alpha\beta}\langle\alpha|\left.K\right|_{\bar{\sigma}}|\beta\rangle a_{\alpha}^{\dagger}a_{\beta},
\]
where $a_{\alpha}^{\dagger},a_{\alpha}$ are the creation and destruction
operators for the complete 1-body basis $|\alpha>$. In the momentum
space basis, there is a natural choice
\begin{equation}
K(\bar{\sigma})=\frac{1}{2V}\sum_{b,\vec{k},\vec{k}'}e^{i(\vec{k}-\vec{k}').\vec{r}}\left(\sqrt{k^{2}+M_{b}[\bar{\sigma}(\vec{r})]^{2}}+\sqrt{k'^{2}+M_{b}[\bar{\sigma}(\vec{r})]^{2}}\right)a_{b\vec{k}}^{\dagger}a_{b\vec{k}'},\label{eq:Kinetic-1}
\end{equation}
where the form has been chosen to guarantee hermiticity and $V$
is the normalisation volume. We also choose 
\begin{equation}
\frac{\partial K}{\partial\sigma}(\bar{\sigma})=\frac{1}{2V}\sum_{b,\vec{k},\vec{k}'}e^{i(\vec{k}-\vec{k}').\vec{r}}\frac{\partial}{\partial\bar{\sigma}}\left(\sqrt{k^{2}+M_{b}[\bar{\sigma}(\vec{r})]^{2}}+\sqrt{k'^{2}+M_{b}[\bar{\sigma}(\vec{r})]^{2}}\right)a_{b\vec{k}}^{\dagger}a_{b\vec{k}'},\label{eq:Kinetic-2}
\end{equation}
with a similar expression for the second derivative. The remaining
ordering ambiguities, arising from products of non-commuting operators, can
be fixed by a normal ordering prescription. 

\subsection{Uniform matter}
\label{UnifMat}

In uniform matter  $\langle K(\bar{\sigma})\rangle$ and its derivatives are independent
of $\vec{r.}$ Using the interacting Fermi gas model we have
\begin{equation}
\langle K_{b}(\bar{\sigma})\rangle=\frac{2}{(2\pi)^{3}}\int_{0}^{k_{F}(b)}d\vec{k}\sqrt{k^{2}+M_{b}^{2}(\bar{\sigma})},\,\,\,\langle K(\bar{\sigma})\rangle=\sum_{b}\langle K_{b}(\bar{\sigma})\rangle,\label{eq:104-1}
\end{equation}
where the Fermi momentum $k_{F}$ is related to the baryonic density by $\rho=gk_{F}^{3}/(6\pi^{2})$
with $g$ being the degeneracy. For the derivatives we have analogous expressions.
The constant expectation value of the sigma field $\bar{\sigma}$
is then determined by the self consistent equation:
\begin{equation}
\frac{dV}{d\sigma}(\bar{\sigma})=-\langle\frac{\partial K}{\partial\sigma}(\bar{\sigma})\rangle,\label{eq:100}
\end{equation}
which is solved numerically.

The fluctuation $\delta\bar{\sigma}$ is given by Eq. ~(\ref{eq:order 1}).
If we define the (constant) effective mass
\begin{equation}
\tilde{m}_{\sigma}^{2}=\frac{d^{2}V}{d\sigma}\left(\bar{\sigma}\right)+\langle\frac{\partial^{2}K}{\partial\sigma^{2}}(\bar{\sigma})\rangle,\label{eq:102}
\end{equation}
the solution is 
\begin{eqnarray}
\delta\sigma(\vec{r}) & = & \int d\vec{r}'\langle\vec{r}\left|\left(-\nabla_{r}^{2}+\tilde{m}_{\sigma}^{2}\right)^{-1}\right|\vec{r}'\rangle\left(-\frac{\partial K}{\partial\sigma}(\bar{\sigma},\vec{r}')+\langle\frac{\partial K}{\partial\sigma}(\bar{\sigma})\rangle\right)\nonumber\\
 & = & \frac{1}{\tilde{m}_{\sigma}^{2}}\langle\frac{\partial K}{\partial\sigma}(\bar{\sigma})\rangle-\int d\vec{r}'\langle\vec{r}\left|\left(-\nabla_{r}^{2}+\tilde{m}_{\sigma}^{2}\right)^{-1}\right|\vec{r}'\rangle\frac{\partial K}{\partial\sigma}(\bar{\sigma},\vec{r}').\label{eq:103}
\end{eqnarray}
Using the expression  for $\delta\sigma$ we get the following expression
for the energy density (normal ordering has been assumed in the correlation
term):
\begin{eqnarray}
\frac{\langle H_{\sigma}\rangle}{V}&=&\langle K(\bar{\sigma})+V(\bar{\sigma}) \rangle+\frac{1}{(2\pi)^{6}}\sum_{h}\int_{0}^{k_{F}(h)}d\vec{k}_{1}d\vec{k}_{2}\dots\nonumber\\
  &&\frac{1}{(\vec{k}_{1}-\vec{k}_{2})^{2}+\tilde{m}_{\sigma}^{2}}
    \frac{\partial}{\partial\bar{\sigma}}\sqrt{k_{1}^{2}+M_{h}^{2}(\bar{\sigma})}\frac{\partial}{\partial\bar{\sigma}}\sqrt{k_{2}^{2}+M_{h}^{2}(\bar{\sigma})} \, .
\label{eq:ED_sigma}
\end{eqnarray}
Finally we must add the contribution of the $\omega$ and $\rho$
exchange, which follow directly from Eqs.~(\ref{eq:H_omega}, \ref{eq:H_rho}).
As usual we define the effective couplings
\[
G_{\sigma}=\frac{g_{\sigma}^{2}}{m_{\sigma}^{2}},\,\,\, G_{\omega}=\frac{g_{\omega}^{2}}{m_{\omega}^{2}},\,\,\, G_{\rho}=\frac{g_{\rho}^{2}}{m_{\rho}^{2}}
\]
and we obtain: 
\begin{eqnarray}
\frac{\langle H_{\omega}\rangle}{V}&=&\frac{G_{\omega}}{2}\left(\sum_{b}w_{b}^{\omega}\rho_{b}\right)^{2}\nonumber\\
&-&G_{\omega}\sum_{b}\left(w_{b}^{\omega}\right)^{2}\frac{1}{(2\pi)^{6}}\int_{0}^{k_{F}(h)}d\vec{k}_{1}d\vec{k}_{2}\frac{m_{\omega}^{2}}{(\vec{k}_{1}-\vec{k}_{2})^{2}+m_{\omega}^{2}}.
\label{eq:ED_omega}
\end{eqnarray}
For the $\rho$ exchange we specify the flavor content by $b=t,m,s$
with $t,m$ being the isospin of the state and its projection and $s$ its
strangeness. Then one has 
\begin{eqnarray}
\frac{\langle H_{\rho}\rangle}{V}&=&\frac{G_{\rho}}{2}\left(\sum_{tms}m\rho_{tms}\right)^{2}
-G_{\rho}\sum_{tmm's}\vec{I}_{mm'}^{t}.\vec{I}_{m'm}^{t}\nonumber\\
&&\frac{1}{(2\pi)^{6}}\int_{0}^{k_{F}(tms)}d\vec{k}_{1}\int_{0}^{k_{F}(tm's)}d\vec{k}_{2}\frac{m_{\rho}^{2}}{(\vec{k}_{1}-\vec{k}_{2})^{2}+m_{\rho}^{2}}\label{eq:ED_rho}
\end{eqnarray}
with $\vec{I}_{mm'}^{t}$ being the isospin matrix which satisfies the
relation: $\vec{I}_{mm'}^{t}.\vec{I}_{m'm}^{t}=\delta_{mm'}m^{2}+t(\delta_{m,m'+1}+\delta_{m',m+1})$
(see the Appendix).

For the pion we obtain
\begin{eqnarray*}
\frac{\langle H_{\pi}\rangle}{V} & = & \frac{1}{2(2\pi)^{6}}\left(\frac{3g_{A}}{5f_{\pi}}\right)^{2}\\
&  & \sum_{tmst'm's'\alpha\mu\nu\sigma\sigma'}
 \langle\sigma,tms|G_{T}^{\mu\alpha}|\sigma't'm's'\rangle\langle\sigma',t'm's'|G_{T}^{\nu\alpha}|\sigma,tms\rangle\\
 &  & \int^{k_{F}(tms)}\int^{k_{F}(t'm's')}d\vec{p}d\vec{p}'\frac{(p-p')^{\mu}(p-p')^{\nu}}{(\vec{p}-\vec{p}')^{2}+m_{\pi}^{2}} \, .
\end{eqnarray*}
Using the explicit expression for the Gamow-Teller matrix element a
little algebra leads to 
\begin{eqnarray}
\frac{\langle H_{\pi}\rangle}{V} & = & \frac{1}{\rho_{B}}\left(\frac{g_{A}}{2f_{\pi}}\right)^{2}\left\{ J_{pp}+4J_{pn}+J_{nn}-\frac{24}{25}\left(J_{\Lambda,\Sigma^{-}}+J_{\Lambda,\Sigma^{0}}+J_{\Lambda,\Sigma^{+}}\right)\right.\nonumber \\
 & + & \left.\frac{16}{25}\left(J_{\Sigma^{-}\Sigma^{-}}+2J_{\Sigma^{-}\Sigma^{0}}+2J_{\Sigma^{+}\Sigma^{0}}+J_{\Sigma^{+}\Sigma^{+}}\right)\right.\nonumber \\
 & + & \left.\frac{1}{25}\left(J_{\Xi^{-}\Xi^{-}}+4J_{\Xi^{-}\Xi^{0}}+J_{\Xi^{0}\Xi^{0}}\right)\right\} 
\label{eq:ED_pion}
\end{eqnarray}
with 
\begin{eqnarray}
J_{bb'} & = & \frac{1}{(2\pi)^{6}}\int^{k_{F}(b)}\int^{k_{F}(b')}d\vec{p}d\vec{p}'\left[1-\frac{m_{\pi}^{2}}{(\vec{p}-\vec{p}')^{2}+m_{\pi}^{2}}\right] \, ,
\label{eq:QMC20}
\end{eqnarray}
where $\rho_{B}=\sum_{b}\rho_{b}$ is the total baryon density. The contact
term in $J_{bb'}$ will be removed, since by hypothesis the bags do
not overlap. In other words we keep only the long range part of the
pion exchange, proportional to $m_{\pi}^{2}$, which is obviously
attractive. In normal matter at saturation it gives a binding of a
few MeV per nucleon. Note that if we were to (erroneously) keep
this contact term the only consequence would be a minor re-adjustment
of the other mesons couplings.

\subsubsection{Cold uniform matter and neutron stars}
\label{ColdNewEoS}

Uniform matter in cold neutron stars is in a generalized beta-equilibrium
(BEM), achieved over a period of time which is extremely long compared with 
time scales typical of the strong and weak interactions (provided no more 
than one unit of strangeness is changed). All baryons of the octet
can be populated by successive weak interactions, regardless of their
strangeness \cite{Glendenning2012}. However the equation of state (EoS) with the octet
$(N,\Lambda,\Sigma,\Xi)$ should be computed in such a way that the
$\Xi$ particles (the cascades) do not appear \emph{unless }the $\Lambda,\Sigma$
hyperons are already present. The reason is that the production of the $\Xi$
from non-strange matter would require a weak interaction with a change
of 2 units of strangeness. The corresponding cross sections are so
small that they cannot appear during the equilibration time of the
star. We implement this feature in the EoS by rejecting the solutions
of the $\beta$ equilibrium equations when they contain a non-zero
density of $\Xi$ while the density of $\Lambda,\Sigma$ zero. By
construction the $\Lambda$ then appears 
\footnote{in the QMC model the $\Sigma$ hyperons 
do not appear in the density range of interest. 
} before the $\Xi$ and, because it is lighter, this tends to increase 
the pressure a little.

This cascade inhibition brings in a new problem because if the chemical
equilibrium equation $\mu_{\Xi^{-}}-\mu_{e}=\mu_{\Lambda}$ is satisfied
at densities below  the $\Lambda$ threshold then it will not be after. So when
the $\Xi^{-}$ creation is again allowed the equilibrium equation
may never be satisfied. However this is a meta-stable situation because
in this case we have $\mu_{\Xi^{-}}-\mu_{e}<\mu_{\Lambda}$ . Therefore
the electron capture reaction $\Lambda+e\to\Xi^{-}+\nu$ is kinematically
allowed and will restore the equilibrium. 

We consider matter formed by baryons of the octet, electrons and negative
muons with respective densities $\{\rho_{b},b=p,n,...\}$ and $\{n_{e},n_{\mu}\}.$
The equilibrium state minimises the total energy density, $\epsilon$, 
under the constraint of baryon number conservation and electric neutrality.
We write:
\begin{equation}
\epsilon=\epsilon_{B}(\rho_{p},...)+\epsilon_{e}(n_{e})+\epsilon_{\mu}(n_{\mu}),\label{eq:beta1}
\end{equation}
where the baryonic contribution is:
\begin{equation}
\epsilon_{B}(\rho_{p},...)=\frac{1}{V}\langle H_{\sigma}+H_{\omega}+H_{\rho}+H_{\pi}\rangle\label{eq:beta2}
\end{equation}
 and is calculated according to Eqs.~(\ref{eq:ED_sigma},\ref{eq:ED_omega},\ref{eq:ED_rho},\ref{eq:ED_pion}).
It is related to the binding energy per particle ${\cal E}$ by:
\begin{equation}
\epsilon_{B}(\rho_{p},...)=\sum_{b}({\cal E}+M_{b})\rho_{b}.\label{eq:beta3}
\end{equation}
In (proton+neutron)
matter another important variable is the symmetry energy, ${\cal S}$, 
which is often defined as the difference between pure neutron and symmetric matter
\begin{equation}
{\cal S}(\rho_{B})={\cal E}\left(\rho_{p}=0,\,\rho_{n}=\rho_{B}\right)-{\cal E}\left(\rho_{p}=\rho_{B}/2,\,\rho_{n}=\rho_{B}/2\right).
\label{eq:beta5}
\end{equation}
For the energy density of the lepton $l$ of mass $m_{l}$ and density
$n_{l}$ we use the free Fermi gas expression:
\begin{equation}
\epsilon_{l}(n_{l})=\frac{2}{(2\pi)^{3}}\int^{k_{F}(l)}d\vec{k}\sqrt{k^{2}+m_{l}^{2}},\,\,\,{\rm with}\, n_{l}=\frac{k_{F}^{3}(l)}{3\pi^{2}}.
\label{eq:beta6}
\end{equation}
The equilibrium condition for a neutral system with baryon density
$\rho_{B}$ writes 
\begin{equation}
\delta\left[\epsilon_{B}(\rho_{p},...)+\epsilon_{e}(n_{e})+\epsilon_{\mu}(n_{\mu})+\lambda(\sum_{b}\rho_{b}-n_{B})+\nu(\sum_{b}\rho_{b}q_{b}-(n_{e}+n_{\mu})\right]=0 \, ,
\label{eq:beta7}
\end{equation}
where $q_{b}$ is the charge of the flavor $b$. The constraints are
implemented through the Lagrange multipliers $\lambda,\nu$, so the
variation in (\ref{eq:beta7}) amounts to independent variations of
the densities together with the variations of $\lambda,\nu$. If one
defines the chemical potentials as 
\begin{equation}
\mu_{b}=\frac{\partial\epsilon_{B}}{\partial\rho_{b}},\,\,\,\mu_{l}=\frac{\partial\epsilon_{l}}{\partial n_{l}}=\sqrt{k_{F}^{2}(l)+m_{l}^{2}},
\label{eq:beta8}
\end{equation}
the equilibrium condition may be written as
\begin{eqnarray}
\mu_{b}+\lambda+\nu q_{b} & = & 0,\,\,\, b=p,...,\label{eq:beta9}\\
\mu_{e}-\nu & = & 0,\label{eq:beta10}\\
\mu_{\mu}-\nu & = & 0,\label{eq:beta11}\\
\sum_{b}\rho_{b}-\rho_{B} & = & 0,\label{eq:beta12}\\
\sum_{fb}\rho_{b}q_{b}-(n_{e}+n_{\mu}) & = & 0.
\label{eq:beta13}
\end{eqnarray}
This is a system of non-linear equations for $\{\rho_{p},...,n_{e},n_{\mu},\lambda,\nu\}$. 
It is usual to eliminate the Lagrange multipliers using Eq.~(\ref{eq:beta10}) 
and Eq.~(\ref{eq:beta9}), with $b=n$. However, for a given value
of $\rho_{B}$, the equilibrium equation (\ref{eq:beta7}) generally
implies that some of the densities vanish and therefore the equations
generated by their variations drops from the system (\ref{eq:beta9}-\ref{eq:beta11})
because there is nothing to vary! In particular substituting $\nu=\mu_{e}$
in Eqs.~(\ref{eq:beta9}) is not valid when the electron disappears
from the system. The equations obtained by this substitution may have
no solution in the deleptonized region, since one has incorrectly forced
$\nu=0$. To correct for this effect, $\nu$ must be restored as an independent variable in this density region. This is technically inconvenient
and we have found it is much simpler to solve blindly 
the full system (\ref{eq:beta9}-\ref{eq:beta13})
for the set $\{\rho_{p},...,n_{e},n_{\mu},\lambda,\nu\}$. The only
simplification which is not dangerous is the elimination of the muon
density in favor of the electron density by combining Eqs.~(\ref{eq:beta10}, \ref{eq:beta11})
to write $\mu_{\mu}=\mu_{e}$, which is solved by 
\begin{equation}
k_{F}(\mu)=\Re\sqrt{k_{F}(e)^{2}+m_{e}^{2}-m_{\mu}^{2}} \, , 
\label{eq:beta14}
\end{equation}
where $\Re$ denotes the real part. This is always correct because
if the electron density vanishes then so does the muon density and
the relation (\ref{eq:beta14}) reduces to $0=0$. 

To solve the system (\ref{eq:beta9}-\ref{eq:beta13}) let us define
the set of relative concentrations (note that the lepton concentrations
are also defined with respect to $\rho_{B}$)
\begin{equation}
Y=\{y_{i}\}=\left\{ \frac{\rho_{p}}{\rho_{B}},\frac{\rho_{n}}{\rho_{B}},\frac{\rho_{\Lambda}}{\rho_{B}},...,\frac{n_{e}}{\rho_{B}},\frac{n_{\mu}}{\rho_{B}}\right\} .
\label{eq:beta15}
\end{equation}
Once the equilibrium solution
$Y(\rho_{B})$ has been found for the desired range of baryon density,
typically $\rho_{B}=0\div1.2 {\rm fm}^{-3}$, it is used to compute
the equilibrium total energy 
density $\epsilon(\rho_{B})=\epsilon(y_{p}\rho_{B},...,y_{\mu}\rho_{B})$
and the corresponding total pressure $P(\rho_{B})$ is computed numerically
as
\begin{equation}
P(\rho_{B})=\rho_{B}^{2}\frac{d}{d\rho_{B}}\frac{\epsilon(\rho_{B})}{\rho_{B}}.
\label{eq:beta16}
\end{equation}

\subsection{Low density expansion for finite nuclei}

We now use the QMC model developed in the previous sections to build
a density functional for Hartree-Fock calculations of finite nuclei.
We write the full QMC hamiltonian as 
\[
H_{QMC}=H_{\sigma}+H_{\omega}+H_{\rho}+H_{\pi}+H_{SO},
\]
where $H_{\sigma},H_{\omega},H_{\rho}$ are given in Eqs.~ (\ref{eq:H_sigma}, \ref{eq:H_omega}, \ref{eq:H_rho})
where the flavor is now restricted to protons and neutrons, which allows
us to set the weights $w_{\sigma},\tilde{w}_{\sigma},w_{\omega}$
to one. In particular, the source of the $\omega$ field becomes the
normal density operator:
\[
D_{\omega}(\vec{r})=\sum_{i}\delta(\vec{r}-\vec{R}_{i})\equiv D(\vec{r})=D_{p}+D_{n}.
\]
The contribution of the pion, $H_{\pi}$, and the spin-orbit term, $H_{SO}$, 
will be added as a perturbation at the end of the derivation. The
reason for this 2-step process is two-fold. First, because of its long-range the pion exchange needs a very specific approximation. Second, 
the spin-orbit term uses the results of the first step.

Thus we first derive the expectation value of $H_{\sigma}+H_{\omega}+H_{\rho}$
in a Slater determinant for $Z$ protons and $N$ neutrons. We assume
that it is obtained by filling the single particle states $\{\phi^{i}(\vec{r},\sigma,m)\}$
up to a Fermi level $F_{m}.$ Here $m=\pm1/2$ is the isospin projection
such that $p\leftrightarrow1/2,n\leftrightarrow-1/2$. In the following
the labels $p,n$ or $\pm1/2$ are used interchangeably according to
the context.

We define the usual C-number densities
\begin{eqnarray}
\rho_{m}(\vec{r}) & = & \langle D_{m}\rangle=\sum_{i\in F_{m}}\sum_{\sigma}\left|\phi^{i}(\vec{r},\sigma,m)\right|^{2},\,\,\,\rho=\rho_{p}+\rho_{n},\label{eq:54}\\
\tau_{m}(\vec{r}) & = & \sum_{i\in F_{m}}\sum_{\sigma}\left|\vec{\nabla}\phi^{i*}(\vec{r},\sigma,m)\right|^{2},\,\,\,\tau=\tau_{p}+\tau_{n},\label{eq:55}\\
\vec{J}_{m} & = & i\,\sum_{i\in F_{m}}\sum_{\sigma\sigma'}\vec{\sigma}_{\sigma'\sigma}\times\left[\vec{\nabla}\phi^{i}(\vec{r},\sigma,m)\right]\phi^{i*}(\vec{r},\sigma',m),\,\,\,\vec{J}=\vec{J}_{p}+\vec{J}_{n} \, ,
\label{eq:56}
\end{eqnarray}
which will be used to write $\langle H\rangle.$

\subsubsection{$\langle H_{\omega}\rangle$, \textmd{\normalsize{}$\langle H_{\rho}\rangle$}}

The expression (\ref{eq:omega-Eq}) for $H_{\omega}$ is not convenient
for our purpose. Instead we use the equivalent expression
\[
H_{\omega}=\frac{1}{2}g_{\omega}\int d\vec{r}\omega(\vec{r})D(\vec{r}),
\]
with $\omega(\vec{r})$ being a solution of Eq.~ (\ref{eq:omega-Eq}), 
which we write in the form
\begin{eqnarray*}
\omega(\vec{r}) & = & \frac{g_{\omega}}{m_{\omega}^{2}}D(\vec{r})+\frac{1}{m_{\omega}^{2}}\nabla_{r}^{2}\omega\\
 & = & \frac{g_{\omega}}{m_{\omega}^{2}}\left(D(\vec{r})+\frac{1}{m_{\omega}^{2}}\nabla_{r}^{2}D(\vec{r})+\frac{1}{m_{\omega}^{4}}\nabla_{r}^{2}\nabla_{r}^{2}D(\vec{r})
+\cdots\right) \, . 
\end{eqnarray*}
Since the range of the $\omega$ exchange is much smaller than the
distance over which the density varies, it is a sensible approximation
to keep the first two terms of the expansion and write:
\[
H_{\omega}\simeq\frac{G_{\omega}}{2}\int d\vec{r}\left(D^{2}(\vec{r})+\frac{1}{m_{\omega}^{2}}D\nabla_{r}^{2}D(\vec{r})\right) \, . 
\]
Using standard techniques we have 
\begin{eqnarray*}
\langle D^{2}\rangle & = & \sum_{mm'}\left(1-\frac{1}{2}\delta_{mm'}\right)\rho_{m}\rho_{m'},\\
\langle D\nabla_{r}^{2}D\rangle & = & \sum_{mm'}\rho_{m}\nabla_{r}^{2}\rho_{m'}-\delta_{mm'}\left[\frac{1}{2}\rho_{m}\nabla_{r}^{2}\rho_{m}+\frac{1}{4}\left(\left(\vec{\nabla}\rho_{m}\right)^{2}+2J_{m}^{2}-4\rho_{m}\tau_{m}\right)\right].
\end{eqnarray*}
We shall follow the common practice of neglecting the $J^{2}$ term 
which vanishes if the spin-orbit partners have the same radial wave
functions. Since we treat the spin-orbit interaction as a first order
perturbation this is consistent. 
Using integration by parts to eliminate~ $\left(\vec{\nabla}\rho_{b}\right)^{2}$ one then obtains
\begin{eqnarray*}
\langle H_{\omega}\rangle & = & \frac{G_{\omega}}{2}\int d\vec{r}\sum_{mm'}\left[\left(1-\frac{1}{2}\delta_{mm'}\right)\rho_{m}\rho_{m'}\right.\\
 &  & \left.+\frac{1}{m_{\omega}^{2}}\left(\rho_{m}\nabla^{2}\rho_{m'}-\frac{1}{4}\delta_{mm'}\left(\rho_{m}\nabla^{2}\rho_{m}-4\rho_{m}\tau_{m}\right)\right)\right] \, . 
\end{eqnarray*}
A similar calculation leads to 
\begin{eqnarray*}
\langle H_{\rho}\rangle & = & \frac{G_{\rho}}{2}\int d\vec{r}\sum_{mm'}\left\{ \left(mm'-\frac{1}{2}\vec{I}_{mm'}.\vec{I}_{m'm}\right)\rho_{m}\rho_{m'}\right.\\
 &  & \left.+\frac{1}{m_{\rho}^{2}}\left(mm'-\frac{1}{4}\vec{I}_{mm'}.\vec{I}_{m'm}\right)\rho_{m}\nabla^{2}\rho_{m'}+\frac{1}{m_{\rho}^{2}}\vec{I}_{mm'}.\vec{I}_{m'm}\rho_{m}\tau_{m'}\right\} \, , 
\end{eqnarray*}
where $\vec{I}_{mm'}.\vec{I}_{m'm}=\delta_{mm'}m^{2}+(\delta_{m,m'+1}+\delta_{m',m+1})/2.$

\subsubsection{$\langle H_{\sigma}\rangle$}

The starting point for this discussion is Eq.~(\ref{eq:<H_sigma>}):
\[
\langle H_{\sigma}\rangle=\int d\vec{r}\,\left[\langle K(\bar{\sigma})+V(\bar{\sigma})\rangle-\frac{1}{2}\bar{\sigma}\left(\langle\frac{\partial }{\partial\sigma}(\bar{\sigma})\rangle+\frac{dV}{d\sigma}\right)+\frac{1}{2}\langle\delta\sigma\frac{\partial K}{\partial\sigma}(\bar{\sigma})\rangle\right] \, .
\]
We first perform the non-relativistic expansion. We define: 
\[
\xi_{op}(\vec{r})=\frac{1}{V}\sum_{\vec{k},\vec{k}'}e^{i(\vec{k}-\vec{k}').\vec{r}}\frac{k^{2}+k'^{2}}{2}a_{\vec{k}}^{\dagger}a_{\vec{k}'},\,\,\,\xi=\langle\xi_{op}\rangle
\]
and since we now limit our considerations to densities of the order
or smaller than the saturation density $\rho_{\rm 0}\simeq 0.15fm^{\rm -3}$
, we can expand the operator $K(\bar{\sigma})$ and its derivatives
to first order in $\xi_{op}$:
\begin{eqnarray}
K(\bar{\sigma}) & \simeq & D(\vec{r})M(\bar{\sigma})+\frac{\xi_{op}(\vec{r})}{2M(\bar{\sigma})}\nonumber \\
\frac{\partial K}{\partial\sigma}(\bar{\sigma}) & \simeq & \frac{\partial M}{\partial\bar{\sigma}}\left(D(\vec{r})-\frac{\xi_{op}(\vec{r})}{2M(\bar{\sigma})}\right),\label{eq:Kine expand}\\
\frac{\partial^{2}K}{\partial\sigma^{2}}(\bar{\sigma}) & \simeq & \frac{\partial^{2}M}{\partial\bar{\sigma}^{2}}\left(D(\vec{r})-\frac{\xi_{op}(\vec{r})}{2M(\bar{\sigma})}\right)+\frac{\xi_{op}(\vec{r})}{M(\bar{\sigma})^{3}}\left(\frac{\partial M}{\partial\bar{\sigma}}\right)^{2} \, .
\nonumber 
\end{eqnarray}

\subsubsection{$\bar{\sigma}$}

We recall that $\bar{\sigma}(\vec{r})$ is a C-number determined by
Eq.~( \ref{eq:order 0}). Using the non-relativistic expansion we can
write it:
\[
-\nabla^{2}\bar{\sigma}+\frac{dV}{d\sigma}\left(\bar{\sigma}\right)=\left(\rho-\frac{\xi(\vec{r})}{2M(\bar{\sigma})}\right)g_{\sigma}\left(1-dg_{\sigma}\bar{\sigma}\right) \, .
\]
We define the zero range solution $\check{\sigma}$ by the equation
\begin{equation}
\frac{dV}{d\sigma}\left(\check{\sigma}\right)=-\rho g_{\sigma}\left(1-dg_{\sigma}\check{\sigma}\right)
\label{eq:zero-range}
\end{equation}
and approximate $\bar{\sigma}$ by retaining only the terms which
are linear in $\nabla^{2}$ or $\xi$. This leads to
\[
\bar{\sigma}\simeq\check{\sigma}+\left(\frac{d^{2}V}{d\sigma^{2}}(\check{\sigma})+\rho dg_{\sigma}^{2}\right)\left(\nabla^{2}\check{\sigma}+g_{\sigma}(1+dg_{\sigma}\check{\sigma})\frac{\xi(\vec{r})}{2M(\check{\sigma})}\right).
\]
In our context it is necessary to have an analytic expression for
$\check{\sigma}$ and we get around the problem by assuming that $\check{\sigma}$
can be represented by a rational fraction. If $\lambda_{3}=\lambda_{4}=0$, 
Eq.~(\ref{eq:zero-range}) has the solution 
\[
g_{\sigma}\check{\sigma}=\frac{G_{\sigma}\rho}{(1+dG_{\sigma}\rho)}.
\]
By analogy we have chosen to write 
\[
g_{\sigma}\check{\sigma}=\frac{G_{\sigma}\rho+\beta\rho^{2}}{1+G_{\sigma}(d+d_{extra})\rho}
\]
and we have fitted the two coefficients $\beta,d_{extra}$ in the range
$\rho=[0.05,0.4]fm^{-3}$.

\subsubsection{$\delta\sigma$}

If we define the effective $\sigma$ mass as
\[
\tilde{m}_{\sigma}^{2}(\bar{\sigma})=\frac{d^{2}V}{d\sigma^{2}}\left(\bar{\sigma}\right)+<\frac{\partial^{2}K}{\partial\sigma^{2}}(\bar{\sigma})>
\]
the equation for $\delta\sigma$, Eq.~(\ref{eq:order 1}), becomes
\[
-\nabla^{2}\delta\sigma+\tilde{m}_{\sigma}^{2}\delta\sigma=-\frac{\partial K}{\partial\sigma}(\bar{\sigma})+\langle\frac{\partial K}{\partial\sigma}(\bar{\sigma})\rangle.
\]
We can solve this equation by again keeping only the terms which are linear
in $\nabla^{2}$ or $\xi$, that is
\begin{equation}
\delta\sigma\simeq\frac{1}{\tilde{m}_{\sigma}^{2}}\left(-\frac{\partial K}{\partial\sigma}(\bar{\sigma})+\langle\frac{\partial K}{\partial\sigma}(\bar{\sigma})\rangle\right)+\frac{1}{\tilde{m}_{\sigma}^{2}}\nabla^{2}\frac{1}{\tilde{m}_{\sigma}^{2}}\left(-\frac{\partial K}{\partial\sigma}(\bar{\sigma})+\langle\frac{\partial K}{\partial\sigma}(\bar{\sigma})\rangle\right).
\label{eq:211}
\end{equation}
The above expression for $\delta\sigma$ still contains terms of higher
order in $\nabla^{2}$ or $\xi$ which we do not include here for simplicity. These higher order terms are dropped at the end of the derivation.

\subsubsection{$\langle H_{\sigma}\rangle$}

If we insert the expression for $\bar{\sigma},\delta\sigma$ in Eq.~
\ref{eq:<H_sigma>} we get, after some algebra:
\begin{equation}
\langle H_{\sigma}^{mean}\rangle=M(\check{\sigma})\rho+\frac{1}{2M(\check{\sigma})}\left[\tau-\frac{1}{2}\nabla^{2}\rho\right]+\frac{1}{2}\left(\vec{\nabla}\check{\sigma}\right)^{2}+V(\check{\sigma})\label{eq:<H_sigma_mean>}
\end{equation}
\begin{eqnarray}
\langle H_{\sigma}^{fluct}\rangle & = & \frac{1}{4\tilde{m}_{\sigma}^{2}(\bar{\sigma})}\left(\frac{\partial M}{\partial\bar{\sigma}}(\bar{\sigma})\right)^{2}\sum_{m}\rho_{m}^{2}\nonumber \\
 & + & \frac{1}{8\tilde{m}_{\sigma}^{2}(\check{\sigma})M^{2}(\check{\sigma})}\left(\frac{\partial M}{\partial\bar{\sigma}}(\check{\sigma})\right)^{2}\left(\sum\rho_{m}\nabla^{2}\rho_{m}-2\sum\rho_{m}\tau_{m}\right)\nonumber \\
 & + & \frac{1}{4\tilde{m}_{\sigma}^{2}(\check{\sigma})}\frac{\partial M}{\partial\bar{\sigma}}(\check{\sigma})\nabla^{2}\left(\frac{1}{\tilde{m}_{\sigma}^{2}}\frac{\partial M}{\partial\bar{\sigma}}(\check{\sigma})\right)\sum_{m}\rho_{m}^{2}\nonumber \\
 & + & \frac{1}{2\tilde{m}_{\sigma}^{2}(\check{\sigma})}\frac{\partial M}{\partial\bar{\sigma}}(\check{\sigma})\vec{\nabla}\left(\frac{1}{\tilde{m}_{\sigma}^{2}}\frac{\partial M}{\partial\bar{\sigma}}(\check{\sigma})\right).\sum_{m}\rho_{m}\vec{\nabla}\rho_{m}\nonumber \\
 & + & \frac{1}{2}\left(\frac{1}{\tilde{m}_{\sigma}^{2}(\check{\sigma})}\frac{\partial M}{\partial\bar{\sigma}}(\check{\sigma})\right)^{2}\times\nonumber\\ 
 &&\left(\frac{1}{2}\sum_{m}\rho_{m}\nabla^{2}\rho_{m}+\frac{1}{4}\sum\left(\vec{\nabla}\rho_{m}\right)^{2}-\sum\rho_{m}\tau_{m}\right) \, ,
\label{eq:<H_sigma_fluct>}
\end{eqnarray}
where we have separated the fluctuation (proportional to $\delta\sigma)$)
from the mean field contribution. 

In our previous work we have used a simplified version of $\langle H_{\sigma}\rangle$
where the contributions proportional to either $\nabla^{2}$ or $\xi$
were truncated to their 2-body parts. This was mostly motivated by
the fact that the Skyrme force, to which we often wished to compare our results,
has such a limitation. Applying these truncations to $\langle H_{\sigma}\rangle=\langle H_{\sigma}^{mean}\rangle+\langle H_{\sigma}^{fluct}\rangle$
leads to the following simple expression: 
\begin{eqnarray}
\langle H_{\sigma}^{simple}\rangle & = & M\rho+\frac{\tau}{2M}+\frac{G_{\sigma}}{2M^{2}}\rho\tau-\left(\frac{G_{\sigma}}{2m_{\sigma}^{2}}+\frac{G_{\sigma}}{4M^{2}}\right)\sum_{m}\rho_{m}\tau_{m}\nonumber \\
 &  & -\left(\frac{G_{\sigma}}{2m_{\sigma}^{2}}+\frac{G_{\sigma}}{4M^{2}}\right)\rho\nabla^{2}\rho+\frac{1}{8}\left(\frac{G_{\sigma}}{m_{\sigma}^{2}}+\frac{G_{\sigma}}{M^{2}}\right)\sum_{m}\rho_{m}\nabla^{2}\rho_{m}\nonumber \\
 &  & -\frac{1}{2}\frac{G_{\sigma}}{1+dG_{\sigma}\rho}\left(\rho^{2}-\frac{1}{(1+dG_{\sigma}\rho)^{2}}\frac{1}{2}\sum_{m}\rho_{m}^{2}\right) \, .
\label{eq:<H_sigma_simple>}
\end{eqnarray}

\subsubsection{Spin-orbit interaction}

We write the spin orbit Hamiltonian starting from Eqs.~(\ref{eq:Vso-magn},\ref{eq:Vso-precess}).
Since the interaction already involves a gradient, one must neglect
any terms in the meson fields containing $\nabla^{2}\rho$ or $\xi$.
Defining 
\[
\vec{{\cal J}}=-\sum_{i}\delta(\vec{r}-\vec{R}_{i})\vec{\sigma}_{i}\times\vec{P}_{i},\,\,\vec{{\cal J}}^{\alpha}=-\sum_{i}\delta(\vec{r}-\vec{R}_{i})\vec{\sigma}_{i}\times\vec{P}_{i}I_{i}^{\alpha}
\]
we obtain
\[
H_{SO}=\int d\vec{r}\left[C^{IS}(\rho)\vec{{\cal J}}.\vec{\nabla}D(\vec{r})+C^{IV}(\rho)\vec{{\cal J}}^{\alpha}.\vec{\nabla}D^{\alpha}(\vec{r})\right] \, , 
\]
where the isoscalar and isovector coefficients are expressed as
\begin{eqnarray*}
C^{IS}(\rho) & = & -\frac{1}{4M^{2}(\sigma)}\left(\frac{\partial}{\partial\rho}M(\check{\sigma})+G_{\omega}\right)+\frac{1}{2MM(\check{\sigma})}\frac{\mu_{IS}(\check{\sigma})}{\mu_{N}}G_{\omega}\\
C^{IV}(\rho) & = & -\frac{G_{\rho}}{4M^{2}(\check{\sigma})}+\frac{G_{\rho}}{2MM(\check{\sigma})}\frac{\mu_{IV}(\check{\sigma})}{\mu_{N}} \, . 
\end{eqnarray*}
As the value of the magnetic moments must  be taken in the local scalar field, 
we have fitted a simple form for this dependence:
\begin{eqnarray*}
\frac{\mu(\check{\sigma})}{\mu(0)} & = & 1+0.547254g_{\sigma}\check{\sigma}-0.0149432\left(g_{\sigma}\check{\sigma}\right)^{2} \, .
\end{eqnarray*}
Finally we get the following expression for the Hartree-Fock expectation value:
\begin{eqnarray*}
\langle H_{SO}\rangle & = & C^{IS}(\rho)\left[\vec{\nabla}.\left(\rho\vec{J}\right)-\frac{3}{2}\left(\rho_{p}\vec{\nabla}.\vec{J}_{p}+\rho_{n}\vec{\nabla}.\vec{J}_{n}\right)-\left(\rho_{p}\vec{\nabla}.\vec{J}_{n}+\rho_{n}\vec{\nabla.}\vec{J}_{p}\right)\right]\\
 & + & C^{IV}\left[\frac{1}{4}\vec{\nabla}.\left[\left(\rho_{p}-\rho_{n}\right)\left(\vec{J}_{p}-\vec{J}_{n}\right)\right]-\frac{3}{8}\left(\rho_{p}\vec{\nabla}.\vec{J}_{p}+\rho_{n}\vec{\nabla}.\vec{J}_{n}\right)\right] \, .
\end{eqnarray*}
Note that if we truncate this expression to 2-body interactions we
recover the expression used in previous work~\cite{Stone:2016qmi}:
\begin{eqnarray*}
\langle H_{SO}^{simple}\rangle & = & \left[\frac{G_{\omega}}{4M^{2}}\left(2\frac{\mu_{Is}}{\mu_{N}}-1\right)+\frac{G_{\sigma}}{4M^{2}}\right]\left[-\frac{3}{2}\left(\rho_{p}\vec{\nabla}.\vec{J}_{p}+\rho_{n}\vec{\nabla}.\vec{J}_{n}\right)-\left(\rho_{p}\vec{\nabla}.\vec{J}_{n}+\rho_{n}\vec{\nabla.}\vec{J}_{p}\right)\right]\\
 & + & \left[\frac{G_{\rho}}{4M^{2}}\left(2\frac{\mu_{IV}}{\mu_{N}}-1\right)\right]\left[-\frac{3}{8}\left(\rho_{p}\vec{\nabla}.\vec{J}_{p}+\rho_{n}\vec{\nabla}.\vec{J}_{n}\right)\right] \, .
\end{eqnarray*}

\subsubsection{Pion in Local Density Approximation (LDA)}
\label{Pion-LDA}
The derivation of the density functional of the QMC model makes extensive
use of the short range approximation which is suggested by the relatively large
masses of the $\sigma,\omega,\rho$ mesons. This, of course, is not
possible for the pion exchange because of the small mass of the pion.
For the latter we use the local density approximation (LDA). Starting
from Eq.~(\ref{eq:H_pion}), written for $p,n$ flavors, one gets
\begin{eqnarray*}
\langle H_{\pi}\rangle & = & \frac{1}{2}\left(\frac{g_{A}}{2f_{\pi}}\right)^{2}\int d\vec{r}d\vec{r}'\sum_{i,j\in F}\left[\phi^{*i}\vec{\sigma}\tau^{\alpha}\phi^{j}\right]_{r}\left[\phi^{*j}\vec{\sigma}\tau^{\alpha}\phi^{i}\right]_{r'}\vec{\nabla_{r}}\vec{\nabla}_{r'}\langle\vec{r}|\left(-\nabla_{r}^{2}+m_{\pi}^{2}\right)^{-1}|\vec{r}'\rangle\\
 & = & \frac{g_{A}^{2}}{8f_{\pi}^{2}}\int\frac{d\vec{q}}{(2\pi)^{3}}d\vec{r}d\vec{r}'\frac{e^{i\vec{q}.(\vec{r}-\vec{r}')}}{q^{2}+m_{\pi}^{2}}Tr\left[\vec{\sigma}.\vec{q}\tau^{\alpha}\rho(\vec{r},\vec{r}')\vec{\sigma}.\vec{q}\tau^{\alpha}\rho(\vec{r'},\vec{r})\right] \, ,
\end{eqnarray*}
where we have defined the non-local density for each flavor
\[
\rho(r,r')=\sum_{i\in F}\phi^{i}(r)\phi^{*i}(r').
\]
After evaluation of the traces one can write
\[
\langle H_{\pi}\rangle=\langle H_{\pi}\rangle_{pp}+\langle H_{\pi}\rangle_{nn}+2\langle H_{\pi}\rangle_{pn}+2\langle H_{\pi}\rangle_{np} \, , 
\]
with
\begin{equation}
\langle H_{\pi}\rangle=\frac{g_{A}^{2}}{8f_{\pi}^{2}}\int\frac{d\vec{q}}{(2\pi)^{3}}d\vec{r}d\vec{r}'e^{i\vec{q}.(\vec{r}-\vec{r}')}\left(1-\frac{m_{\pi}^{2}}{q^{2}+m_{\pi}^{2}}\right)\rho_{m}(\vec{r},\vec{r}')\rho_{n}(\vec{r'},\vec{r}) \, .
\label{eq:<H_pion_LDA>}
\end{equation}
The LDA amounts to computing $\rho(\vec{r},\vec{r}')$ in the Fermi
gas approximation with a Fermi momentum evaluated using the local density
at $\vec{R}=(\vec{r}+\vec{r}')/2$. This gives 
\begin{equation}
\rho(\vec{r},\vec{r}')=\frac{3\pi^{2}}{k_{F}^{3}(\vec{R})}\int_{0}^{k_{F}(\vec{R})}\frac{d\vec{k}}{(2\pi)^{3}}\rho(\vec{R})e^{i\vec{k}.(\vec{r}-\vec{r}')}.
\label{eq:LDA}
\end{equation}
We have also tried the improved LDA proposed in Ref.~\cite{Gebremariam2010}
but we found that it leads to instabilities in the Hartree-Fock self-consistent
calculation without obvious improvements. Using (\ref{eq:LDA}) in
(\ref{eq:<H_pion_LDA>}), where we have removed the contact piece, leads
to: 
\begin{eqnarray*}
\langle H_{\pi}\rangle_{mn} & = & -\frac{9m_{\pi}^{2}g_{A}^{2}}{32f_{\pi}^{2}}\int d\vec{R}\frac{\rho_{m}(\vec{R})\rho_{n}(\vec{R})}{k_{Fm}^{3}(\vec{R})k_{Fn}^{3}(\vec{R})}\\
&&\int_{0}^{k_{Fm}(\vec{R})}dk\int_{0}^{k_{Fn}(\vec{R})}dk'\int_{-1}^{1}du\frac{k^{2}k^{\prime2}}{k^{2}+k^{\prime2}-2kk'u+m_{\pi}^{2}} \, .
\end{eqnarray*}

\subsubsection{Hartree-Fock equations}

Our derivation of the QMC density functional is now complete. From
it we can derive the Hartree-Fock equations for the single particle states $\phi^{i}(\vec{r},\sigma,m)$
\[
\sum_{\sigma'}\left[\delta_{\sigma\sigma'}\left(-\vec{\nabla}.\frac{1}{2M_{eff}(m)}\vec{\nabla}+U(m)\right)+i\vec{W}(m).\vec{\sigma}_{\sigma\sigma'}\times\vec{\nabla}\right]\phi^{i}(\vec{r},\sigma',m)=e_{i}\phi^{i}(\vec{r},\sigma,m)
\]
with 
\[
U(m)=\frac{\delta\langle H_{QMC}\rangle}{\delta\rho_{m}(\vec{r})}-M,\,\,\,\frac{1}{2M_{eff}(f)}=\frac{\delta\langle H_{QMC}\rangle}{\delta\tau_{m}(\vec{r})},\,\,\: W^{\alpha}(m)=\frac{\delta\langle H_{QMC}\rangle}{\delta J_{m}^{\alpha}(\vec{r})}.
\]
We do not write the expressions of the Hartree-Fock potentials here as they are far
too long. In practice they are passed directly from Mathematica to
the Fortran code.

\section{Applications}
\label{applications}

Since its introduction in 1988, the basic idea of the QMC model has attracted wide spread attention and has been used, at various levels of complexity, to model properties of hadronic matter under different conditions. In this section we wish to give selected examples of the application of the QMC theory in the past, as well as report the latest results obtained with the full QMC-II model introduced in Section~\ref{qmc}. There is no space in this review to discuss technical differences between 
the individual variants of the model used in the past and we refer the reader to the original papers. However, we wish to stress the versatility of the model, even in a somewhat simplified form, 
as well as its prospects for the future.

\subsection{Nuclear matter}
\label{nuclearmatter}

One of the main advantages of the QMC model is that different phases of hadronic matter, from very low to very high baryon densities and temperatures, can be described within the same underlying framework. Although the QMC model shares some similarities with QHD \cite{Serot1997} and the Walecka-type models \cite{Serot1986}, there are significant differences. 
Most importantly, in QMC the internal structure of the nucleon is introduced explicitly. In 
addition, the effective nucleon mass lies in the range 0.7 to 0.8 of the free nucleon mass (which agrees with results derived from non-relativistic analysis of scattering of neutrons from lead nuclei \cite{Johnson1987}) and is higher than the effective nucleon mass in the Walecka model. Also, at finite temperature at fixed baryon density, the nucleon mass always decreases with temperature in QHD-type models while it increases in QMC. However, the lack of solid experimental and or observational data prevents selection of a preferred model and one is just left  with a description of differences between individual predictions.

In the QMC model, infinite nuclear matter and finite nuclei are intimately related, in other words, the model is constructed in such a way that it predicts the properties of the two systems consistently. As explained later in Section~\ref{finite}, nuclear matter properties are always a starting point in the process of determining the parameters of the model Hamiltonian for finite nuclei.  We therefore refer the reader to references in Section~\ref{finite}, covering earlier results and the exploitation of infinite nuclear matter properties in calibrating the QMC model parameters. In this subsection we will focus on the use of QMC predictions in modeling the dense matter appearing in compact objects. 

\subsubsection{Phase transitions and instabilities at sub-saturation density}

One of the interesting areas of application of the QMC model is the transitional region between the inner crust and outer core of a cold neutron star (at densities just below the nuclear saturation density $\rho_0$). The phenomena that are predicted to occur in this region include instabilities, formation of droplets and/ or appearance of the ``pasta'' phase both at zero and finite temperatures.

 Krein et al. \cite{Krein2000d} used the QMC model to study droplet formation at $T=0$ during the liquid to gas phase transition in cold asymmetric nuclear matter. The critical density and proton fraction for the phase transition were determined in the mean field approximation. Droplet properties were calculated in the Thomas-Fermi approximation. The electromagnetic field was explicitly included and its effects on droplet properties studied. The results were compared with those obtained with the NL1 parametrization of the non-linear Walecka model and the similarities and differences discussed.

The earliest application of the QMC model at finite temperature was reported by Song and Su \cite{Song1995}. The resulting EoS was applied to discuss the liquid-gas phase transition in nuclear matter below the saturation density. The calculated critical temperature for the transition and temperature dependence of the effective mass were compared with those given by the Walecka and other related models.

The equation of state of warm (up to $T$ = 100 MeV) asymmetric nuclear matter in the QMC model and mechanical and chemical instabilities were studied as a function of density and isospin asymmetry \cite{Panda2003}. The binodal section, essential in the study of the liquid-gas phase transition, was also constructed and discussed. The main results for the equation of state were compared with two common parametrizations used in the non-linear Walecka model and the differences were outlined. The mean meson effective fields were determined from the minimization of the thermodynamical potential, and the temperature dependent effective bag radius was calculated from the minimization of the effective mass of the nucleon mass of the bag. The thermal contributions of the quarks, which are absent in the Walecka model, was dominant and led to a rise of the effective nucleon mass at finite temperatures. This was contrary to the results presented in \cite{Song1995}, where temperature was introduced only at the hadron level, and therefore the behavior of the effective mass with temperature was equivalent to the results of Walecka-type models. The effective radius of the nucleon bag was found to shrink with increasing temperature.

Thermodynamical  instabilities for both cold symmetric and asymmetric matter within the QMC model, with (QMC$\delta$) and without (QMC) the inclusion of the isovector-scalar $\delta$ meson were studied by Santos et al. \cite{Santos2009}. 
The model parameters were adjusted to constraints on the slope parameter of the nuclear symmetry energy at saturation density. The spinodal surfaces and predictions of the instability regions 
obtained in the QMC and QMC$\delta$ models were compared with results of mean field relativistic models and discussed.

Grams {\it et al.}~\cite{Grams2017} studied the pasta phases in low density regions of nuclear and neutron star matter within the context of the QMC model. Fixed proton fractions as well as nuclear matter in $\beta$-equilibrium at zero temperature were considered. It has been shown that the existence of the pasta phases depends on choice of the surface tension coefficient and the influence of the nuclear pasta on some neutron star properties was examined.

\subsubsection{The EoS of high density matter in neutron stars and supernovae}

Some applications of the QMC model in building the EoS of neutron stars have utilized simplified
expressions for the energy of the static MIT bag, representing the baryons, and the effective mass of the nucleon taken equal to the energy of the bag. The meson fields were treated as classical fields in a mean field approximation \cite{Panda2004b, Panda2004a, Panda2010a, Panda2012}. The quark matter considered in some of these models was treated using the EoS from  \cite{Farhi1984} and related references. 

Panda {\it et al.}~\cite{Panda2004b} built an EoS for a hybrid neutron star with mixed hadron and quark phases. The QMC model was used for the hadron matter, including the possibility of creation of hyperons. Two possibilities were considered for the quark matter phase, namely, the unpaired quark phase (UQM), described by a simple MIT bag, and the color-flavor locked (CFL) phase in which quarks of all three colors and flavors are allowed to pair and form a superconducting phase. The bag constant $B^{\rm 1/4}$ was varied between 180 - 211 MeV (190 - 211) for QMC+UQM (QMC-CFL)  systems and the highest neutron star mass of 1.85 $M_\odot $ was predicted for QMC+UQM with $B^{\rm 1/4}$ =211 MeV. It is interesting to note that the $u$, $d$ and $s$ quarks appeared in the QMC+UQM matter at densities as low as about twice $\rho_{\rm 0}$ before, or competing with, the appearance of hyperons, depending on the value of the bag constant (see Fig. 4 in Ref.~\cite{Panda2004b} ).  
This work was followed by Ref.~\cite{Panda2004a}, where the effect of trapped neutrinos in 
a hybrid star was studied. It was found that a neutrino-rich star would have larger maximum baryonic mass than a neutrino poor star.This effect would lead to low-mass black hole formation during the leptonization period.

Neutrino-free stellar matter and matter with trapped neutrinos at fixed temperatures and with the entropy of the order of 1 or 2 Boltzmann units per baryon was studied in the QMC model by Panda {\it et al.}~\cite{Panda2010a}. A new prescription for the calculation of the baryon effective masses in terms of the free energy was used. 
Comparing the results with those obtained from the non-linear
Walecka model, smaller strangeness and neutrino fractions were predicted within the QMC model. As a consequence, it was suggested that the QMC model might have a smaller window of 
metastability for conversion into a low-mass blackhole during cooling.

The QMC model has been adjusted to provide a soft symmetry energy
density dependence at large densities in \cite{Panda2012}. 
The hyperon-meson couplings were chose QMC n according
to experimental values of the hyperon nuclear matter potentials, and possible uncertainties were considered.  The hyperon content and the mass/radius curves for the
families of stars obtained within the model were discussed. It has been shown that a softer symmetry energy gives rise to stars with less hyperons, smaller radii  and larger masses. It was found that the hyperon-meson couplings may also have a strong effect on the mass of a star \cite{Panda2012}.

A fully self-consistent, relativistic, approach based on the theory detailed in Sections~\ref{UnifMat} (except for the inhibition of cascade production which will be explained in the following paragraph) was used by Stone {\it et al.}~\cite{RikovskaStone:2006ta} to construct the EoS and to calculate key properties of high density matter and cold, slowly rotating neutron stars. The full baryon octet was included in the calculation. The QMC EoS provided cold neutron star models with maximum mass in the range 1.9 - 2.1 $M_\odot$, with central density less than six times nuclear saturation density  and offered a consistent description of the stellar mass up to this density limit three years before their observation \cite{Demorest:2010bx,Antoniadis:2013pzd}. 

In contrast with other models, the QMC EoS  predicted no hyperon contribution at densities lower than 3$\rho_{\rm 0}$, for matter in $\beta$-equilibrium. At higher densities, $\Lambda$ and $\Xi^{\rm 0,-}$ hyperons were present, with consequent lowering of the maximum mass as compared with matter containing only nucleons, electrons and muons but still reaching the maximum gravitational mass $M_{\rm g}$=1.99 $M_\odot$. A key reason for the higher maximum mass possible within the QMC model is the automatic inclusion of repulsive three-body forces between hyperons and nucleons as well as hyperons and hyperons. These are a direct consequence of the scalar polarizability of the composite baryons and their prediction requires no new parameters. We also note that the model predicts the absence of lighter $\Sigma^{\rm \pm,0}$ hyperons, which is at variance with the results of most earlier models. This may be understood as consequence of including the color hyperfine interaction in the response of the quark bag to the nuclear scalar field. This finding was later observed and discussed by other models (see e.g. \cite{Schaffner-Bielich2010a}). We summarize the main results of Ref.~\cite{RikovskaStone:2006ta} in Table~\ref{2007model}. In addition,  conditions related to the direct URCA process were explored and the parameters relevant to slow rotation, namely the moment of inertia and the period of rotation, were investigated. 
\begin{table}
\centering{}\caption{\label{2007model}Selected properties of neutron star models predicted
in \cite{RikovskaStone:2006ta}. $\rho_{{\rm c}},R_{c},M_{g},v_{s}$
are central baryon density, radius, gravitational mass and the speed
of light for the maximum and $1.4M_{\odot}$neutron star models, respectively.
FQMC600 (700) are models with full baryon octet and the mass of the
$\sigma$ meson 600 (700) MeV. NQMC600(700) are model for nucleon-only
matter. Data for N-QMCx EoS are not shown because the central density
of all $1.4M_{\odot}$ stars is predicted to be below the threshold
for the appearance of hyperons.}
\begin{tabular}{|c|c|c|c|c||c|c|c|c|}
\hline
Model & $\rho_{c}$ & $R_{c}$ & $M_{g}$ & $v_{s}$ & $\rho_{c}$ & $R_{c}$ & $M_{g}$ & $v_{s}$\tabularnewline
 & $({\rm fm}^{-3})$ & (km) & $(M_{\odot})$ & (c) & $({\rm fm}^{-3})$ & (km) & $(M_{\odot})$ & (c)\tabularnewline
\hline 
\hline 
F-QMC600 & 0.81 & 12.45 & 1.99 & 0.65 & 0.39 & 12.94 & 1.4 & 0.58\tabularnewline
\hline 
F-QMC700 & 0.82 & 12.38 & 1.98 & 0.65 & 0.39 & 12.88 & 1.4 & 0.58\tabularnewline
\hline 
N-QMC600 & 0.96 & 11.38 & 2.22 & 0.84 & \multicolumn{1}{c}{} & \multicolumn{1}{c}{} & \multicolumn{1}{c}{} & \multicolumn{1}{c}{}\tabularnewline
\cline{1-5} 
N-QMC700 & 0.96 & 11.34 & 2.21 & 0.84 & \multicolumn{1}{c}{} & \multicolumn{1}{c}{} & \multicolumn{1}{c}{} & \multicolumn{1}{c}{}\tabularnewline
\cline{1-5} 
\end{tabular}
\end{table}
One important recent development that was not accounted for in earlier work is the suppression 
of $\Xi$ hyperons until $\Lambda$ hyperons are allowed by the chemical equilibrium equations. 
This was explained earlier, in Section~\ref{UnifMat}.  In the follow-up study the instability, arising at 
and slightly above the threshold density for appearance of $\Lambda$ hyperons at about $\sim$ 0.55 fm$^{\rm -3}$ 
was treated by interpolation of the EoS connecting the two regions, nucleon-only and 
nucleon+$\Lambda$ +$\Xi^{\rm 0,-}$, with electrons and muons present in both regions. 
\begin{figure}
\centering{}\includegraphics[height=6cm,width=9cm]{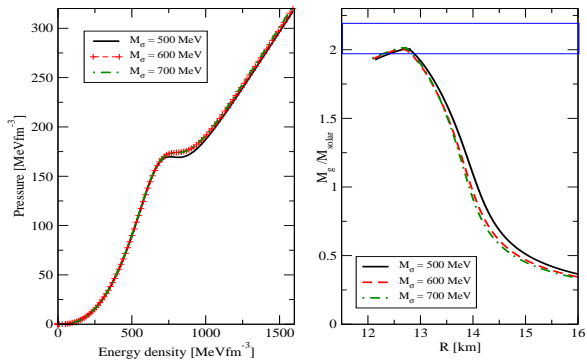}
\caption{Left: The EoS yielding neutron star models with maximum mass 2.005, 2.006 an 2.014 M$_{\odot}$ for M$_\sigma$ - 500, 600 and 700 MeV, respectively. Right: Gravitational mass vs radius for these models. The (blue) rectangle depicts the current observational limits on maximum mass of cold non-rotating neutron stars.}
\label{review_mass}
\end{figure}
We note that the $\beta$-equilibrium is recovered at densities lower than the central baryon density of the neutron star with 
maximum mass. This instability, not studied before, may have important consequences for the neutron star physics. 
We feel that it should be included in this review in order to focus on one of the future developments of 
consequences of the QMC model that should not be overlooked.  The preliminary results are illustrated 
in Fig.~\ref{review_mass}, illustrating the interpolated EoS and the mass-radius curves for three values of 
the mass of the $\sigma$ meson 500, 600 and 700 MeV. The predicted maximum mass of the neutron star 
in each model is within the latest limits set by Rezzolla {\it et al.}~\cite{Rezzolla2017a}, derived from the 
observation of gravitational waves from neutron star mergers under the condition that the 
product of a merger will collapse to a black hole.

\subsubsection{The Fock Term}

In versions of the QMC model for nuclear matter which utilize the Hartree-Fock technique, 
the effect of the Fock term has been examined by several authors \cite{Krein1999,Massot2012,Miyatsu2012,Katayama2012,Miyatsu2013b,Whittenbury2014}. 
Detailed discussion of these approaches to the exchange term in the QMC Hamiltonian 
may be found in Ref.~\cite{Whittenbury2014}.

Whittenbury {\it et al.}~\cite{Whittenbury2014} included the full vertex structure of the exchange term, containing not only the Dirac vector term, as was done in Ref.~\cite{RikovskaStone:2006ta}, but also the Pauli tensor term. These terms, already in QMC cluded within the QMC model by Krein {\it et al.}~\cite{Krein1999} for symmetric nuclear matter, were generalized by evaluating the full exchange terms for all octet baryons and adding them, as additional contributions, to the energy density. A consequence of this increased level of sophistication is that, if one insists on using the hyperon couplings predicted in the simple QMC model, i.e. with no coupling to the strange quarks, that the $\Lambda$  hyperon is no longer bound. It is remarkable that in the absence of the Pauli Fock terms, the model predicted realistic $\Lambda$  binding energies and, at the same time, realistic $\Sigma$  repulsion in matter~\cite{RikovskaStone:2006ta}. It turns out that the additional repulsion associated with the Fock term is not adequately compensated and the agreement is lost. The magnitude of the needed change by artificially modifying the $\sigma$ couplings for the hyperons to match the empirical observations was studied in detail. This procedure was designed to serve as a guidance in the future development of the model.

\subsubsection{Chiral QMC models of nuclear matter}

Chiral versions of the QMC (CQMC) model have been utilized by several authors, mainly to explore different phases of neutron star crust and interior and to study exotic formations of hybrid and quark stars. Although differing somewhat in the techniques used in the original QMC model, the basic ideas are preserved.

Miyatsu {\it et al.}~\cite{Miyatsu2013} used a CQMC model and applied it to uniform nuclear matter within the relativistic Hartree-Fock approximation. The EoS was constructed considering the full baryon octet in the core region and nuclei in the Thomas-Fermi approximation in the crust. They found that only the $\Xi^{\rm -}$ hyperon appeared in the neutron star core and the maximum mass was predicted to be 1.95 $M_\odot$.

The CQMC model, based on the $SU(3)$ linear $\sigma$ model with the vacuum pressure and 
vector meson exchange included,  was used to describe the properties of compact stars made 
of cold pure quark matter \cite{Zacchi2015}. Variation of the vector coupling constant, 
the mass of constituent quarks in vacuum, which fixes the scalar meson coupling constant, 
and the vacuum constant which does not effect the scalar field but just shifts the energy density 
at a given pressure, were studied. It was found that a stable pure quark configuration with 
maximum mass $\sim$ 2 $M_\odot$ can be realized with a reasonable set of parameters. 

The same model has been applied to hybrid stars \cite{Zacchi2016}, assuming that the pure quark 
core is surrounded by a a crust of hadronic matter. Taking a density dependent hadronic EoS and a 
density dependent chiral quark matter EoS, the transition between the two phases was studied and 
conditions for the appearance of \textit{twin} stars were discused. This work was further extended \cite{Zacchi2017,Zacchi2017a} to finding a new stable solutions of the Tolman-Oppenheimer-Volkoff 
equations for quark stars. These new solutions were found to exhibit two stable branches in the 
mass-radius relation, allowing for twin stars; i.e., two stable quark star solutions with the same 
mass, but distinctly different radii. These solutions are compatible with causality, the stability 
conditions of dense matter and the 2$M_\odot$ pulsar mass constraint.

The CQMC has been  investigated for the two- and three-flavor case extended by contributions of vector mesons under conditions encountered in core-collapse supernova matter \cite{Beisitzer2014}. Typical temperature ranges, densities and electron fractions, as found in core-collapse supernova simulations, were  studied by implementing charge neutrality and local $\beta$-equilibrium with respect to weak interactions. 
Within this framework, the resulting phase diagram and equation of state (EoS) were analysed  and the impact of 
the undetermined parameters of the model were investigated.

\subsubsection{Boson condensates}

In the previous sections only fermionic species have been considered to be present in hadronic matter. However, it may be possible
that boson condensates could play an important role in understanding behaviour of of the matter under extreme conditions, 
especially in connection with the role of strangeness in the cores of neutron stars.
 
Tsushima {\it et al.}~\cite{Tsushima1998a} investigated the properties of the kaon, K, and anti-kaon, $\bar{K}$, in nuclear 
medium using the QMC model.  Employing a constituent quark-antiquark MIT bag model picture, they calculated their 
excitation energies in a nuclear medium at zero momentum within mean field approximation. The scalar and the vector 
mesons were assumed to couple
directly to the non-strange quarks and anti-quarks in the K and $\bar{K}$ mesons. It was demonstrated that the $\rho$ meson induces
different mean field potentials for each member of the iso-doublets, K and $\bar{K}$, when they are embedded in asymmetric
nuclear matter. Furthermore, it was also shown that this $\rho$ meson potential is repulsive for the K$^-$ meson in matter with a
neutron excess, which rendered  K$^-$  condensation less likely to occur.

However, Menezes {\it et al.}~\cite{Menezes2005} studied properties of neutron stars, consisting of a crust of hadrons and an internal part of hadrons and kaon condensate within the QMC model. In the hadron phase nucleon-only stars as well as stars with hyperons were considered. The maximum mass of the neutron star was predicted to be 2.02, 2.05, 1.98, 1.94  $M_\odot$ for np, np+kaon, np+hyperons and np+hyperons+kaon systems, respectively. The kaon optical potentials at saturation density  were of the order of -24 MeV for K$^+$, almost independent of the bag radius,  K$^-$ exhibited a strong dependence, varying from -123 MeV  at $R_B$ =0.6 fm to -98 MeV at   $R_B$ =1.0 fm. In the model with hyperons, $\Lambda$, $\Sigma^{\pm}$ and K$^-$ appeared for baryon density below 1.2 fm$^{\rm -3}$.  
Without hyperons, K$^-$ appeared at baryon density $\sim$ 0.5 fm$^{\rm -3}$.

Proto-neutron star properties were studied within a modified version of the QMC model that incorporates $\omega-\rho$ mixing 
plus kaon condensed matter at finite temperature \cite{Panda2014}. 
Fixed entropy as well as trapped neutrinos were taken into account. The results were compared with those obtained with the GM1
parametrization of the non-linear Walecka model for similar values of the symmetry energy slope. Contrary to
GM1, the QMC model predicted the formation of low mass black holes during cooling. It was shown
that the evolution of the proto-neutron star may include the melting of the kaon condensate, driven by the neutrino
diffusion, followed by the formation of a second condensate after cooling. The signature of this process
could be a neutrino signal followed by a gamma-ray burst. They showed that both models, the modified QMC and the 
non-linear Walecka model,  could, in general, describe very massive stars.

\subsection{Finite nuclei}
\label{finite}

In this section we survey the development of the QMC model for investigation of properties of finite nuclei. The QMC concept does not allow readjustment of the parameters to improve agreement with experiment, but requires further development of the model itself through successive stages. At each of the stages, there is only one parameter set to work with, in contrast to other density dependent effective forces, such as the Skyrme force with a multiple parameter sets employed for the same Hamiltonian in attempts to improve agreement with particular selections of experimental and/or observational data.  It is instructive to follow the path towards the QMC current status. A full account of the status of the QMC model prior to 2007 can be found in the review of Saito {\it et al.}~\cite{Saito2007}. This review covers later years, while making 
reference to earlier models where necessary for continuity.

\subsubsection{Doubly closed shell nuclei}

The first application of the QMC model to finite nuclei was reported by Guichon {\it et al.}~\cite{Guichon1996}, following on the 
original formulation \cite{Guichon1988, Guichon1989} and the further developments in 
Refs.~ \cite{Saito1994d, Saito1995b,Saito1994e,Saito1995}. The equation of motion of an MIT bag (the nucleon) in an 
external field was solved self-consistently and the correction of the centre-of-mass motion was added correctly for the first time.  
Having explicitly approached the nuclear matter problem,  one can solve for the properties of finite nuclei \textit{without explicit reference to the internal structure of the nucleon}. Both non-relativistic and relativistic version of the QMC model were presented. The latter calculation of nuclear matter properties has shown that the model leads naturally to a generalisation of QHD \cite{Serot1986} with a density dependent scalar coupling. The non-relativistic model, with the spin-orbit interaction included, has been applied to predict the charge density distribution in $^{16}$O and $^{40}$Ca, as well as the single-particle proton and neutron states in these nuclei, 
in promising agreement with experiment.
Other groups also worked on applications of the original QMC model \cite{Guichon1988} to finite nuclei. These applications have been restrained to even-even closed shell nuclei, typically $^{16}$O, $^{40}$Ca, $^{48}$Ca,  $^{90}$Zr and $^{208}$Pb ( see e.g. \cite{Fleck1990,Blunden1996a,Saito1997a,Shen2000}).

The relation between the quark structure of the nucleon and  effective, many-body nuclear forces was further developed by Guichon and Thomas \cite{Guichon2004}.  They studied the relation between the effective force derived from the QMC model and the Skyrme force approach. 
A many-body effective QMC Hamiltonian, which led naturally to the appearance of many-body forces, was constructed, considering the zero-range limit of the model. The appearance of the many-body forces was a natural consequence of the introduction of the scalar polarizability in the QMC approach. A comparison of the Hamiltonian with that  of a Skyrme effective force yielded similarities, allowing a very satisfactory interpretation of the Skyrme force which had been proposed on purely empirical grounds. However, the QMC and the Skyrme 
approaches differ in important details, as discussed in Section~\ref{summary}.  The QMC coupling constants $G_\sigma$, $G_\omega$and $G_\rho$  were fixed to produce energy per particle $E/A$ = -15.85 MeV, saturation density $\rho_{\rm 0}$ = 0.16 fm$^{\rm -3}$ and the symmetry energy coefficient $a_{\rm 4}$ = 30 MeV. The fixed parameters of the QMC model, the bag radius and the mass of the $\sigma$ meson, which is not well known experimentally, were taken as $R_B$ = 0.8 fm and $m_\sigma$ = 600 MeV. The Skyrme parameters $t_{\rm 0}$, $x_0$,  $t_3$,
 $5t_2-9t_1$, the effective mass $M_{eff}/M$=$[1+(3t_1+5t_2)M\rho_0/8]^{-1}$ and the strength of the spin-orbit coupling, $W_0$, 
were expressed in terms of  $G_\sigma$, $G_\omega$ and $G_\rho$ and shown to be close to the values obtained for the SIII Skyrme parameterization \cite{Beiner1975a}. 

\subsubsection{Nuclei outside closed shells}

The study of the physical origin of density dependent forces of the Skyrme type was further pursued by 
Guichon {\it et al.}~\cite{Guichon:2006er}. New approximations were introduced to the model, in order to allow calculation of properties of high density uniform matter in the same framework. For finite nuclei this leads to density dependent forces, which compared well to the SkM* Skyrme parameterization \cite{Bartel1982}. The effective interaction, derived from QMC \cite{Guichon2004} has been applied, within the Hartree-Fock-Bogoliyubov (HFB) approach, to doubly closed shell nuclei as well as to the properties of nuclei far from stability. The calculations were performed for the doubly magic nuclei, $^{\rm 16}$O, $^{\rm 40}$Ca, $^{\rm 48}$Ca and $^{\rm 208}$Pb. Reasonable 
agreement was found between experiment and the calculated ground state binding energies, charge rms radii and spin-orbit splittings.  Proton and neutron density distributions from the QMC model were compared to those obtained with the SLy4 Skyrme force \cite{Chabanat1998}  and found to be very similar. Going away from the closed shell, a density dependent contact pairing interaction was included and the two-neutron drip-line predicted for Ni and Zr nuclei. In addition, the shell quenching, predicted by the QMC-HFB model, was demonstrated using the variation of S$_{\rm 2N}$ across N = 28  for two extreme values of proton numbers, namely Z = 32 (proton drip-line region) and Z = 14 (neutron drip-line region). One thus finds that S$_{\rm 2N}$ changes by about 8 MeV for Z = 32 and by about 2-3 MeV for Z=14. This strong shell quenching is very close to that obtained in the Skyrme-HFB calculations (see Fig. 15 of \cite{Chabanat1998}).

These results have been later confirmed and extended by Wang {\it et al.}~\cite{Wang2011}, who used the method in \cite{Guichon:2006er} to include the spin-exchange which, through the Fock exchange term, affects both finite nuclei and nuclear matter. In the QMC model this effect leads to a non-linear density-dependent isovector channel and changes the density-dependent behavior of the symmetry energy. They derived a Skyrme parameterization $Sqmc$ depending on  $t_0-t_3,\,x_0-x_3,\,W_0$ and $\alpha$ which was successfully applied to ground state binding energies of even-even Sn nuclei and proton and neutron charge distributions in $^{\rm 208}$Pb. Wang {\it et al.} also looked into the proton and neutron effective mass at saturation as a function of (N-Z)/A, as well as the density dependence of the symmetry energy. 
However, in the investigation of the latter they took the scalar polarizability $d$ as a variable parameter. 
\begin{figure}
\centering{}\includegraphics[height=7cm,width=7cm]{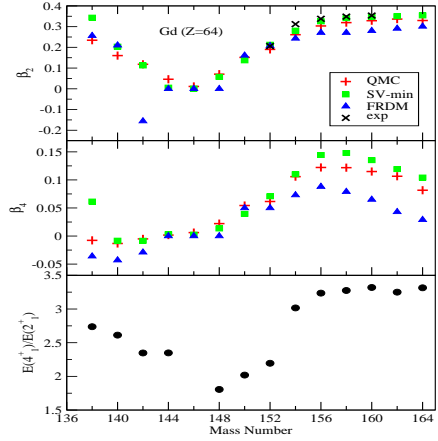}
\caption{Ground state quadrupole deformation parameters $\beta_{\rm 2}$ and  $\beta_{\rm 4}$ of even-even Gd isotopes as calculated with QMC-I and SVmin, compared with the Finite-Range-Droplet-Model (FRDM) \cite{Moeller2016}. The experimental ratios of energies of the first 2$^+$ and 4$^+$ excited states are added as further evidence for appearance of axially deformed shapes.}
\label{fig:shapes}
\end{figure}

A more comprehensive application of the QMC model  has been performed by Stone {\it et al.}~\cite{Stone:2016qmi}, using the same version of the model, labeled QMC-I,  as in Ref.~\cite{Guichon:2006er}. A broad range of ground state properties of even-even spherical and deformed, axially symmetric 
nuclei, as well as nuclei with octupole deformation were studied across the periodic table in the non-relativistic 
Hartree-Fock + BCS framework. For the first time, the QMC parameters $G_\sigma$, $G_\omega$ and $G_\rho$ were not fixed to 
just one set of symmetric nuclear matter saturation properties, as in the previous studies. Because these properties are known only with some uncertainty, it was argued that all combinations of the QMC coupling constants consistent with
$-17 {\rm MeV}<E/A<-15 {\rm MeV}$, $0.14 {\rm fm}^{-3}<\rho_0<0.18 {\rm fm}^{-3}$ for the saturation energy and density and $28 {\rm MeV}<a_4<34 {\rm MeV} $, $L>20 {\rm MeV}$, and $250 {\rm MeV}<K_0<350 \rm{MeV}$  for the symmetry energy coefficient,
its slope and the incompressibility at the saturation should be considered.  The search for
combinations of  $G_\sigma$, $G_\omega$ and $G_\rho$ satisfying these
conditions as a function of $m_\sigma$ was performed on a mesh $9.0 {\rm fm}^2<G_\sigma<14.0 {\rm fm}^2$,  $6.0 {\rm fm}^2<G_\omega ,G_\rho<14.0{\rm fm}^2$ with a step size of $0.5 fm^{\rm 2}$  and $650 MeV <m_\sigma< 750 MeV$
with a step size of 25 MeV.  The result was a well defined region in
the parameter space within which the parameter set, best describing finite
nuclei, was to be sought. The large number of allowed combinations
of obviously correlated parameters rendered a direct search for a unique
set, describing nuclear matter and finite nuclei equally successfully, impractical and a more efficient
approach needed to be adopted.
\begin{table}
\begin{centering}
\caption{\label{tab:PRL} Results of the fit to experimental data in the set
selected by Kl\"upfel et al.\cite{Klupfel2009},
yielding the parameters of the QMC-I model. Equivalent results for the
Skyrme SV-min force \cite{Klupfel2009} are added
for comparison (top part). The rms errors were obtained following
the procedure described in \cite{Klupfel2009}. In
addition, rms errors (no weighting), quantifying the agreement between
calculated and experimental ground state binding energies of selected
$N=Z$ nuclei, $N=Z\pm2$ , 4 mirror nuclei, and selected spherical
and deformed nuclei with $\left|N-Z\right|$ ranging from 2 to 60,
not included in the fit of parameters are shown (bottom part). See
text for more explanation.}
\begin{tabular}{|c|c|c|}
\hline 
data & \multicolumn{1}{c}{rms} & error\%\tabularnewline
\hline 
\hline 
 & QMC & SV-min\tabularnewline
\hline 
fit nuclei: &  & \tabularnewline
binding energies & 0.36 & 0.24\tabularnewline
diffraction radii & 1.62 & 0.91\tabularnewline
surface thickness & 10.9 & 2.9\tabularnewline
rms radii & 0.71 & 0.52\tabularnewline
pairing gap (n) & 57.6 & 17.6\tabularnewline
pairing gap (p) & 25.3 & 15.5\tabularnewline
ls splitting (p) & 15.8 & 18.5\tabularnewline
ls splitting (n) & 20.3 & 16.3\tabularnewline
\hline 
\hline 
super heavy nuclei & 0.1 & 0.3\tabularnewline
N=Z nuclei & 1.17 & 0.75\tabularnewline
mirror nuclei & 1.50 & 1.00\tabularnewline
other & 0.35 & 0.26\tabularnewline
\hline 
\end{tabular}
\par\end{centering}
\end{table}

The QMC EDF (energy density functional)  was incorporated to a Hartree-Fock+BCS code \textit{skyax}~[P.G.Reinhard Personal communication] and the final QMC parameters (QMC-I further on) obtained by fitting to a data set consisted of selected binding energies, $rms$ and diffraction charge radii, surface thickness of the charge distributions, the proton and neutron pairing gaps, and the spin-orbit splitting and energies of single-particle proton and neutron states, distributed across the nuclear chart. The fitting protocol developed by Kl\"upfel
{\it et al.}~\cite{Klupfel2009} was used.  The results are summarized in Table~\ref{tab:PRL}, adopted from Ref.~\cite{Stone:2016qmi}, together with data obtained using a Skyrme EDF with the SV-min Skyrme force \cite{Klupfel2009}.
This work demonstrated the potential of the QMC EDF to predict not only binding energies of even-even nuclear ground states but also their shapes, as illustrated in Fig.\ref{fig:shapes} for nuclear chains from neutron deficit to neutron heavy nuclei, including shell closures. Particularly good agreement between theory and experiment was found for super-heavy nuclei.

Despite the encouraging results obtained in Ref.~\cite{Stone:2016qmi} there were some deficiencies of the QMC EDF which needed improvement. In particular, the incompressibility, $K =340\pm3$ MeV, and the slope of the symmetry energy, $L = 23\pm4$, were somewhat out of the generally expected range. As discussed in \cite{Stone:2016qmi}, the contribution of a long-range Yukawa single pion exchange was expected to lower the incompressibility from 340 MeV to close to 300 MeV. This effect was tested in the next version of the model, QMC-I-$\pi$, ~\cite{Stone2017d}, was used to study even-even superheavy nuclei in the region 96 < Z < 136 and 118 < N < 320. 
\begin{figure}
\centering{}\includegraphics[width=10.0cm]{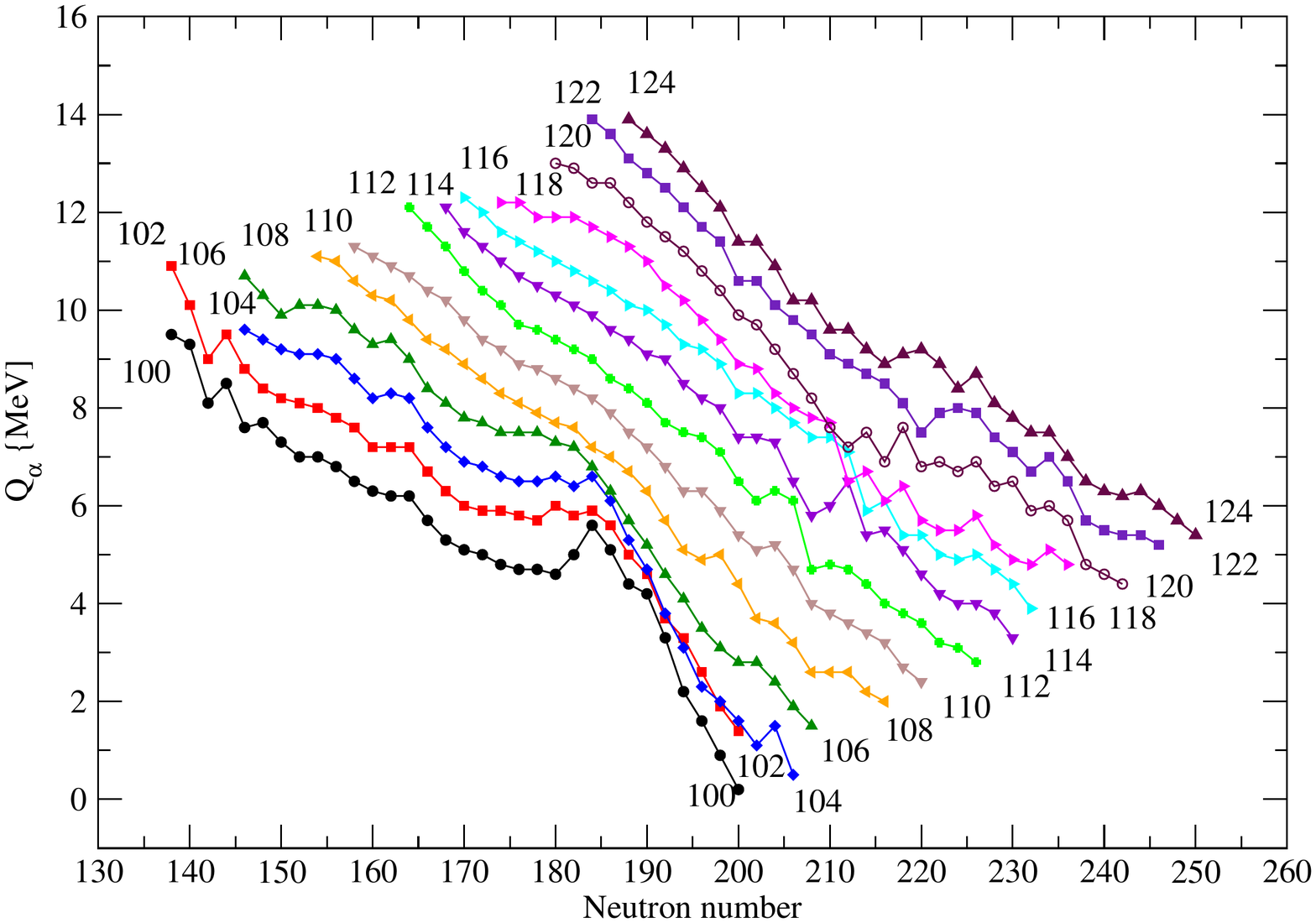}
\caption{Values of the $\alpha$ particle separation energy, $Q_\alpha$, calculated in QMC-I-$\pi$ for isotopes with 100 < $Z$ < 120 in the region of  neutron numbers 138 < $N$ < 252.}
\label{fusion_fig5}
\end{figure}
\begin{figure}
\centering{}\includegraphics[width=10.0cm]{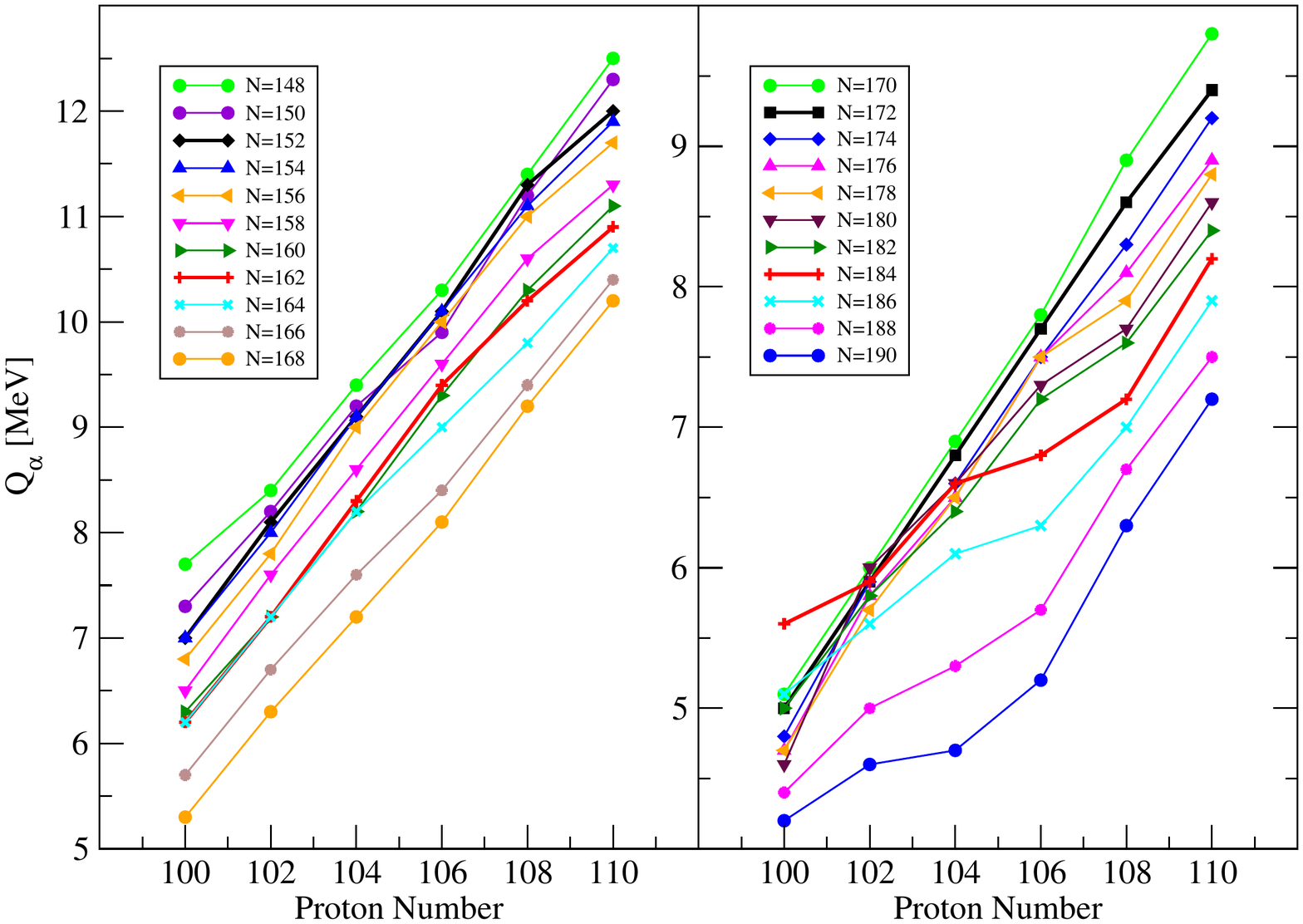}
\caption{The same as Fig.\ref{fusion_fig5}  but for isotones with 148 < $N$ < 190  in the region of proton numbers 100 < $Z$ < 110. The $N$=152, 162, 172 and 184 systems, corresponding to the suggested shell closures in different models, are highlighted by thicker lines in both panels.}
\label{fusion_fig6}
\end{figure}

The QMC EDF was constructed in the same way as in Ref.~\cite{Stone:2016qmi} but included the contribution of the single pion exchange. The parameters of the model were obtained using the experimental data set by Kl\"{u}pfel et al. \cite{Klupfel2009} and the fitting package POUNDERS \cite{Kortelainen2010a,Wild2015}.  The volume pairing in the BCS approximation was adopted, with proton and neutron pairing strength fitted to data in \cite{Klupfel2009}. It is important to note that the addition of the explicit pion exchange in the model  did not increase the number of  parameters beyond the four used in \cite{Stone:2016qmi}, but its addition was reflected in slight changes (less than 5\%)  from the values reported in there. The new parameter set is compatible with nuclear matter properties $E_0$  =-15.8 MeV, $\rho_{\rm 0}$ = 0.153 fm $^{-3}$, $K_{\rm 0}$ = 319 MeV, $a_4$ = 30 MeV and $L$ = 27 MeV. 

As a feature not explored in the previous version (QMC-I) model \cite{Stone:2016qmi}, predictions for $Q_\alpha$ values were reported for the first time. Knowledge of $\alpha$ decay life-times is crucial for predicting properties of the $\alpha$ decay chains of superheavy elements, used for the experimental detection of new elements and their isotopes. Thus the calculation of $Q_\alpha$ as close to reality as possible is vital for planning experiments. The $\alpha$-decay life-times are exponential functions of the energy release, $Q_\alpha$,  in the decay, which, in turn, depends on the mass difference between the parent and daughter states. This means that while the absolute values of the nuclear masses are not needed very precisely in this context, the differences are essential. 

The proton number dependence of $Q_\alpha$ obtained by Stone {et al.}~\cite{Stone2017d} is illustrated in Fig.~\ref{fusion_fig5} and the neutron number dependence in Fig.~\ref{fusion_fig6}. The QMC findings were not compared to results of other model predictions in the literature (e.g. \cite{Heenen2015,Carlsson2016,Jachimowicz2014,Wang2016} and references therein) but there is one important result which has not been observed in any of the other models; namely that the (weak) effects of the N = 152 and 162 shell closures disappear in nuclei with Z > 108, while the effects are enhanced for  N$\sim$180. Thus, in the QMC model a smooth neutron number dependence of  $Q_\alpha$ for N<200 for all elements with Z up to 124 is predicted, not showing any effects of shell structure. Some variations may be indicated for higher N but for these no systematic conclusions could be drawn. 

The outcome of the QMC-I-$\pi$ model indicated that there is a subtle interplay between proton and neutron degrees of freedom in developing regions of nuclei with increased $\alpha$-decay half life. As already discussed  in the literature (e.g. \cite{Bender2001e}), it seems likely that the sharp shell closures and shape changes observed in lighter nuclei, will instead be manifest as smoother patterns around the expected ``shell closures''. These patterns have their origin in the competition between the Coulomb repulsion and surface tension of the large nuclear systems in which the single-particle structure is only one of the critical ingredients. 

While the fundamental feature of the QMC model is that it should describe nuclear matter equally as well as finite nuclei, it has become clear that the addition of the single-pion exchange does not yield desirable values of $K$ and $L$.  Therefore a new version of the QMC EDF (QMC-II) has been developed.

First the $\sigma$ field potential energy
now includes a cubic and quartic terms. The motivation is that it
allows the contribution of the $\sigma$ exchange in the $t$ channel
to the scalar polarizability, something which is beyond the bag model
calculation used until now. So the potential energy is written as
\[
V(\sigma)=\frac{1}{2}m_{\sigma}^{2}\sigma^{2}+\sum_{N=3,4}\frac{1}{N!}\lambda_{N}\left(g_{\sigma}\sigma\right)^{N} \, ,
\]
with the new parameters $\lambda_{3},\lambda_{4}.$ We note that the
cubic term is enough to generate the $t$-channel pole but we include
the quartic term with $\lambda_{4}>0$ to ensure energy positivity.
In practice we can set it to any small value so that it is not a new
free parameter. We constrain the value of $\lambda_{3}$ so that the
extra polarizability it generates is of the same order of magnitude
as the one coming from the bag model ($s$-channel). This amounts
to setting $\lambda_{3} \leq 0.04 {\rm fm}^{-1}$. The effect of the cubic
coupling is moderate at normal nuclear density but it can nevertheless
reduce the incompressibility by 10\%.

The second improvement concerns the approximation scheme which has
been used to derive the QMC energy density.  In the previous derivations it was assumed that
the finite range terms (involving the Laplacian of the density) and
the velocity terms should only be of two-body nature. This simplification
was motivated by the fact the Skyrme interaction, which often
serves as a point of comparison, satisfies this condition. However there
is no obvious reason to support such an assumption, except simplicity.
On the contrary, one might suspect that it could lead to neglecting some important
effects. For instance, the effective mass of the $\sigma$ meson is actually
truncated to its bare value by such an approximation. Similarly the
spin-orbit potential depends on the effective nucleon mass in a relativistic
theory and this effect is lost in the simple approximation scheme.

In the improved low density expansion of the model the energy density contains at most
Laplacian or squared gradient of the density, but with coefficients
which can be arbitrary functions of the density. The main benefit of this improvement
is that the model now predicts the slope of the symmetry energy $L$ around 60 - 70 MeV,
consistent with the other model predictions, see e.g. Ref.~\cite{Horowitz2014}.

Another development  in this latest version has been a more sophisticated search procedure to find the QMC parameters, as well as a 
determination of their realistic errors..  The search performed in the QMC-I-$\pi$ version has been reversed, in that the 
mesh of allowed values of nuclear matter parameters at saturation was defined first 
as -17 < $E/A$ < -15 MeV, 0.15 < $\rho_{\rm 0}$ < 0.17 fm$^{\rm -3}$ and  29 < $a_4$ < 32 MeV and 260 < $K_0$< 320 MeV. 
The search through the mesh led to several thousand values for $G_\sigma$, $G_\omega$and $G_\rho$ and $\lambda_{\rm 3}$ 
combinations. The distribution of each of the parameters has been statistically analysed and trimmed by quantile cuts. 
The selected values were then used as starting points for the POUNDERS fitting of finite nuclei according to the protocol in \cite{Klupfel2009}.
The error of each of the parameters, including their correlations, has been determined from the covariance matrix as in \cite{Kortelainen2010a}.
The final parameters, giving the best fit to finite nuclei, were compatible with nuclear matter parameters $E/A$ = -15.9 MeV, $\rho_{\rm 0}$=0.15 fm$^{\rm -3}$, $K$ = 280 MeV and $L$=68.6 MeV.
\begin{figure}
\centering{}\includegraphics[height=7cm,width=8cm]{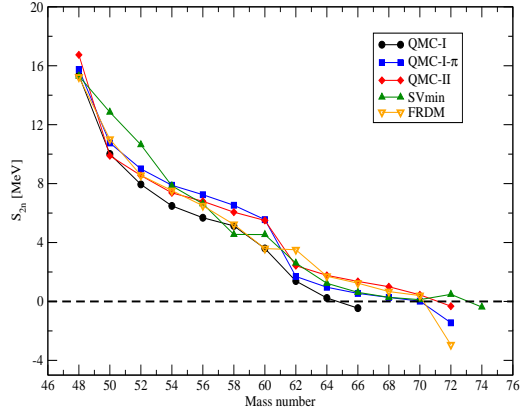}
\caption{Two-neutron separation energies in Ca isotopes calculated in the QMC-II model as a function of mass number. Data from QMC-I, QMC-I-$\pi$, Skyrme SVmin and FRDM are added for a comparison. }
\label{s2n_review}
\end{figure}
The performance of this QMC-II model is illustrated on selected properties of even-even Ca isotopes, which have been in focus of interest, among others, to \textit{ab initio} models \cite{Hagen2012b,Hagen2015}.

In Fig.\ref{s2n_review} the two-neutron separation energies are shown as a function of neutron number, identifying the two neutron drip-line. It is interesting to see that the inclusion of the single-pion exchange in the QMC-I model dramatically increases the drip line from N=66 to N=74, in line with predictions of the SVmin Skyrme and FRDM \cite{Moeller2016}.  Another point to notice is that the large decrease of S$_{\rm 2N}$ across the N=28 shell closure is exhibited by all models in Fig.\ref{s2n_review} but the Skyrme SVmin.
\begin{figure}
\centering{}\includegraphics[height=7cm,width=8cm]{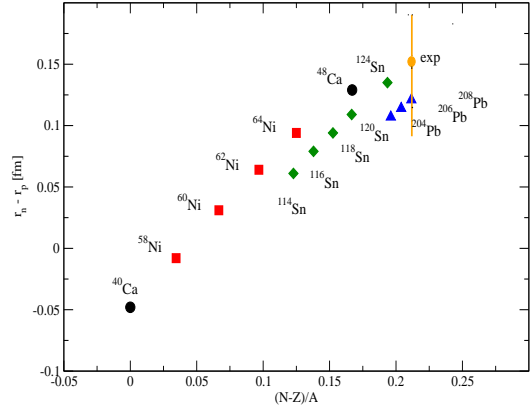}
\caption{Neutron skin in  even-even Ca, Ni, Sn and Pb nuclei as calculated in QMC-II : Predictions for isotopes accessible to experiments [D.P. Watts et. al., MAMI Proposal Nr MAMI-A2-15-09]. Preliminary result for $^{40}$Ca is ~-0.04fm [D.P.Wats, Personal Communication]. The experimental point for $^{208}$Pb is taken from \cite{Tarbert2014}. }

\label{skin_review}
\end{figure}
There has been a lot of interest in charge and matter radii of Ca isotopes \cite{GarciaRuiz2016}. In particular, a large increase in the $rms$ charge radius of $^{\rm 52}$Ca as compared to $^{\rm 48}$Ca as been measured. The charge distribution in $^{\rm 48}$Ca being almost the same as that of $^{\rm 40}$Ca has been a challenge to theorists for decades. As illustrated by Hagen et al. \cite{Hagen2015}, even the latest theories fail to reproduce the value of  $\delta\langle r^{\rm 2}\rangle ^{\rm 48,52}$, as reported in \cite{GarciaRuiz2016}. However, if we compare the experimental  $\langle r^2 \rangle$ between $^{\rm 40}$Ca and  $^{\rm 52}$Ca \cite{GarciaRuiz2016} to results of the QMC-II model, we find full 
agreement, as demonstrated in Table~\ref{Ca48}.
\begin{table}
\begin{centering}
\caption{\label{Ca48}Some properties of $^{\rm 48}$Ca. Results of NNLO$_{sat}$ were taken from \cite{Hagen2015} and Fig.3 in \cite{GarciaRuiz2016}.  All entries are in fm.}
\begin{tabular}{|c||c|c|c|c|c|}
\hline 
Model & $R_{p}$ & $R_{skin}$ & $R_{n}$ & $\delta\langle r^{2}\rangle^{48,52}$ & $\delta\langle r^{2}\rangle^{40,52}$\tabularnewline
\hline 
\hline 
QMC-I & 3.420 & 0.129 & 3.549 & 0.330 & 0.682\tabularnewline
\hline 
QMC-I-$\pi$ & 3.427 & 0.123 & 3.550 & 0.345 & 0.715\tabularnewline
\hline 
QMC-II & 3.388 & 0.16 & 3.554 & 0.330 & 0.538\tabularnewline
\hline 
SV-min & 3.449 & 0.184 & 3.633 & 0.192 & 0.347\tabularnewline
\hline 
$\rm{NNLO_{sat}}$  & 3.405 & 0.128 & 3.533 & 0.3 & \tabularnewline
\hline 
Exp. &  &  &  & 0.530(5)(17) & 0.531(5)(15)\tabularnewline
\hline 
\end{tabular}
\par\end{centering}
\end{table}
Finally, we compare the predictions of the QMC-II model for the neutron skin in even-even Ca, Ni, Sn and Pb isotopes with experiment [D.P. Watts et. al., MAMI Proposal Nr MAMI-A2-15-09]  for isotopes potentially accessible to experiment. It is interesting to note that the QMC model predicts a negative neutron skin in $^{40}$Ca, in agreement with experiment [D.P. Watts, Personal communication] and some \textit{ab initio} models [G. Hagen et al., Personal communication].
 
\subsection{Hypernuclei}
\label{hypernuclei}

The QMC model has been applied to  $\Lambda$ hypernuclei \cite{Tsushima1997,Tsushima1998b} in the past. Most recently,
Guichon {\it et al.}~\cite{Guichon:2008zz}, included the effect of the mean scalar field in-medium on the hyperfine interaction between quarks, arising from one-gluon exchange, self-consistently. The calculations properties of $\Lambda$ and $\Xi$ hypernuclei were of comparable quality to earlier QMC results, without the additional parameter needed there. Even more significantly, the additional repulsion associated with the increased hyperfine interaction in-medium completely changes the predictions for $\Sigma$ hypernuclei. Whereas in the earlier work they were bound by an amount similar to $\Lambda$ hypernuclei, the new model predicts them unbound, in qualitative agreement with the experimental absence of such states, except for very light cases ($^{\rm 4}$He and $^{\rm 9}$Be) \cite{Harada2006g,Harada2015o}. The equivalent non-relativistic potential felt by the $\Sigma$ hyperons has been found repulsive inside the nuclear interior and weakly attractive in the nuclear surface, as suggested by the analysis of $\Sigma$-atoms.

One of the major successes of the QMC model concerns the well-known fact that the spin-orbit force for a $\Lambda$-hyperon is 
exceptionally small. The argument is straightforward. From the explanation in Section~\ref{BaryonInExternal}, we know that the microscopic origin of the spin-orbit force has two sources, a magnetic piece associated with the variation of the mean vector fields across the hadronic volume and a purely geometric piece arising from Thomas precession. Because the light quarks in the $\Lambda$ have zero spin, the first component makes no contribution in this case. As the $\Lambda$ has isospin zero, the total spin-orbit force for the $\Lambda$ therefore comes entirely from the Thomas precession associated with acceleration produced by the $\sigma$ and $\omega$ mean fields. Since these almost exactly cancel, the result is indeed very small.

Shyam et al. \cite{Shyam2012} used the QMC model to predict production cross sections for cascade hypernuclei $^{\rm 12}_\Xi{^-}$Be and $^{\rm 28}_\Xi{^-}$Mg in the (K$^-$, K$^+$) reaction on $^{\rm12}$C and $^{\rm28}$Si targets. This was the first time that the quark degrees of freedom had been explicitly invoked in the description of such reactions. 

\subsection{Symmetry violation}

We observe that when the QMC model is supplemented by the addition of an isovector scalar field, the in-medium variation of the neutron-proton mass difference can provide a semi-quantitative explanation~\cite{Saito1994c} of the Okamoto-Nolen-Schiffer anomaly \cite{Okamoto1964, Nolen1969a,Miller1990e}. 

The effect of nuclear structure variation in super-allowed Fermi beta-decay has been examined in Ref.~\cite{Guichon:2011gc}, including quark mass differences. While the $u$-$d$ mass difference can produce a correction to the value of $V_{ud}$ needed to ensure unitarity of the 
Cabibbo-Kobayashi-Maskawa matrix which is almost significant (i.e., at a level of a few times $10^{-5}$) compared with modern tests using super-allowed Fermi beta-decay of nuclear iso-triplets, the additional correction associated with the modification of nucleon structure in-medium was shown to be an order of magnitude smaller.

Recently Guichon and Thomas~\cite{Guichon:2017gbe} explored another aspect of $\beta$-decay, namely the variation of the 
rate of semi-leptonic decay of the $\Lambda$-hyperon in a hypernucleus, resulting from the change in the  wave function of the 
non-strange quark in the final nucleon. Both the strangeness changing vector and axial weak charges were shown to be reduced 
by up to 10\% at nuclear matter density. Much of the interest in this test of the fundamental hypothesis of the QMC model lies 
in the potential for measuring this effect in a future experiment  at J-PARC. 

\subsection{Alternative nucleon models}
\label{alt}

While the vast majority of the investigations of the importance of a change of hadron structure in-medium have been based upon 
the MIT bag model, its limitations have led to developments involving more sophisticated models of hadron structure. For example, 
if one wishes to investigate the EMC effect, the universally used static cavity approximation creates a number of technical problems. 
While it has been possible to reformulate the problem in such a way as to preserve energy-momentum 
conservation~\cite{Signal:1989yc,Schreiber:1991tc},
which is essential to maintain the correct support of the parton distribution functions, one would like to do better.
This was a key motivation for the reformulation by Bentz and Thomas~\cite{Bentz:2001vc} of the QMC model, 
using the NJL model for hadron structure instead of the MIT bag. Such an approach is much more complicated as one must 
self-consistently solve the Faddeev equation for nucleon structure in medium. On the other hand, because the NJL model 
is covariant, no approximations are required in order to calculate the deep inelastic structure functions~\cite{Mineo:2003vc}. 
The application of this model to the EMC effect will be briefly reviewed in Section~\ref{beyond}, below. 

The NJL model also allows a more reliable calculation of nucleon form factors at high momentum transfer~\cite{Cloet:2014rja}. 
Hence this formulation of the underlying ideas of the QMC model has also been employed recently to investigate the 
expected change in the electromagnetic form factors of a bound nucleon~\cite{Cloet:2015tha}. 
This was also be discussed briefly in Section~\ref{beyond}.

Yet another application of the fundamental idea of the QMC model was recently reported by Bohr {\it et al.}~\cite{Bohr2016}, who developed a model for symmetric and asymmetric nuclear matter based on a QMC model in which the MIT bag was replaced by a Bogoliyubov model [P. N. Bogoliubov, Ann. Inst. Henri Poincare 8 (1968) 163] of the nucleon, in which the quarks are confined by a linearly rising potential. The model predicts, at saturation density, the compressibility $K$ = 335.17 MeV, the quark effective mass $m^*_{\rm q}$ = 238.5 MeV and the effective nucleon mass $M^*$ = 0.76 $M$, where $M$ is the nucleon mass in vacuum. Neutron star masses above two solar masses were obtained.

This section demonstrated the impressively wide versatility of the simple idea that the underlying mechanism of hadronic interactions is the meson exchange between constituent quarks in their interior. There are many techniques and variations of the QMC model in the literature. To access which of the many models are close to reality, more experimental and observational data are needed and the fingerprints of various phenomena need to be determined. 

\section{Signatures of in-medium changes of hadron structure}
\label{beyond}

Until now we have been concerned with the QMC model as an underlying theory for the properties of dense matter and finite nuclei. 
The fundamental feature of the model, namely that what occupies the shell model orbitals are clusters of quarks with nucleon quantum 
numbers but modified internal structure, tends to be hidden in those calculations. For example, the EDF derived within the QMC model 
has a different functional dependence on nuclear density but otherwise resembles traditional Skyrme forces. Nevertheless one must 
ultimately establish the reality or otherwise of these modifications. To this end we briefly review ideas aimed at measuring changes in 
the electromagnetic form factors and structure functions of bound nucleons.

\subsection{The EMC effect}

At the time of its discovery by the European Muon Collaboration~\cite{Aubert:1983xm}, 
the fact that nuclear structure functions differed from those of free 
nucleons in a fundamental way, which could not be understood in terms of Fermi motion, 
was totally unexpected -- see Ref.~\cite{Geesaman:1995yd,Norton:2003cb} for reviews. 
For our purposes, this effect, which is known as the EMC effect, shows an unambiguous loss of momentum from the 
valence quarks in a nucleus. There is as yet no consensus concerning the origin of the effect.

Not surprisingly, the first scientific question asked after the QMC model was first proposed by Guichon was whether or not it was 
capable of explaining some part of the EMC effect. After all, the modification of the valence quark wave functions 
in a bound nucleon is fundamental to the model. Within a year it had indeed been shown that the key 
features of the EMC effect in the valence region could be understood within 
the QMC model~\cite{Thomas:1989vt,Saito:1992rm}. Of course, as explained earlier these calculations were based 
upon the static cavity approximation for the MIT bag model, with its limitations for such problems. 

It took a further 
decade, after the development by Bentz and Thomas of an extension of the QMC model based upon the covariant, 
chiral symmetric, NJL model for hadron structure (as described above in subsect.~\ref{alt}), 
that more sophisticated calculations of the EMC effect became possible~\cite{Cloet:2005rt,Cloet:2006bq}. This 
work established that indeed the predictions of the QMC model were in quantitative agreement with data on the 
EMC effect across the periodic table. Even more important, the model predicted an EMC effect for polarized 
nuclear structure functions that was at least as large. This prediction will be tested in the near future at Jefferson 
Lab following its 12 GeV upgrade. The importance of this measurement is that another proposal for 
the origin of the EMC effect, namely that it arises through nucleons far off-mass-shell because they experience short 
range correlations~\cite{Hen:2016kwk,Frankfurt:1985cv}, does not seem to be compatible with a 
significant polarized EMC effect.

Another prediction based upon the QMC model, which should also be tested at Jefferson Lab, is that in a 
nucleus with $N \neq Z$ there will be what is termed an isovector EMC effect~\cite{Cloet:2009qs}, 
with valence $u$ quarks loosing momentum to the valence $d$ quarks. This can be investigated with 
parity violating deep inelastic scattering~\cite{Cloet:2012td} or, even better, through charged current 
weak interactions at a future electron-ion collider~\cite{Thomas:2009ei}.

\subsection{Elastic electromagnetic form factors}

\subsubsection{$G_E/G_M$}

The extraction of relatively small changes in the electromagnetic form factors of a bound nucleon is experimentally 
very difficult. This realization led Strauch and collaborators to consider a sophisticated new approach. They used the 
features of the Jefferson Lab facility to measure the ratio of two ratios, thus removing or minimizing many potential 
systematic errors. In particular, they used measurements of the longitudinal and transverse recoil polarization of a proton scattered quasi-elastically from $^4$He, as well as a free proton. In this way they could very accurately extract 
the ratio of the electric to magnetic form factors of a proton in $^4$He to the same ratio for a free proton. After careful 
corrections for the effects of distortion of the outgoing proton wave function 
in the He case~\cite{Udias:2000ig,Udias:1999tm}, their 
results~\cite{Strauch:2002wu,Malace:2008gf} were in excellent agreement with the predictions, 
made almost a decade earlier~\cite{Lu:1997mu,Thomas:1998eu,Lu:1998tn}, based upon the QMC 
model. The experimental values clearly disagreed with the hypothesis that there was no modification of the 
bound proton structure. 

For the present the interpretation of this beautiful experiment has been muddied by a suggestion that there might 
be a large, spin-dependent charge exchange correction to this experiment~\cite{Schiavilla:2004xa}. Until this 
rather unlikely possibility is tested experimentally we cannot draw a firm conclusion from the analysis of 
this experiment.

\subsubsection{Coulomb sum rule}

Around the time of the discovery of the EMC effect, Meziani and collaborators published measurements of the 
response functions of a number of nuclei measured in 
electron scattering~\cite{Meziani:1984is,Morgenstern:2001jt} which 
suggested a significant change in the electric form factor of bound protons. Unlike the EMC effect, this work was 
met with widespread criticism from a community convinced that nucleon structure should not change in-medium.
Nevertheless a number of theorists did investigate the potential phenomenological 
consequences of nucleon swelling in medium.
Investigations within the QMC model were first carried out by Saito and collaborators~\cite{Saito:1999bv} and more 
recently using the NJL model for the structure of the nucleon~\cite{Cloet:2015tha}. In both cases the calculations 
support the idea that there is a suppression of the longitudinal response as a result of the modification of the 
electric form factor of the bound proton. 

The Coulomb sum rule was proposed long ago as a means to access short-range correlations in 
nuclei~\cite{McVoy:1962zz}. Experimentally it can be constructed by integrating the longitudinal response 
function over the energy transfer. As the 3-momentum transfer, $|\vec{q}|$ becomes large, one expects 
the Coulomb sum rule to tend to unity, at least in a non-relativistic theory. This is indeed observed in the 
sophisticated Monte-Carlo calculations of the response function of $^{12}$C by 
Lovato {\it et al.}~\cite{Lovato:2013cua}. However, the recent work of Cl\"oet {\it et al.}~\cite{Cloet:2015tha}, 
which included the effect of the change of proton structure (described by the NJL model) induced by the scalar 
mean field in-medium, as well as RPA correlations and relativistic corrections, predicted a suppression of the 
Coulomb sum rule for Pb at high momentum transfer in excess of 40\%. More than half of this suppression 
arose from the change in structure, while relativity accounted for most of the rest. A sophisticated re-measurement 
of this sum rule, which was carried out at Jefferson Lab, is in the final stages of analysis and the results 
are awaited with great anticipation. 

\section{Summary}
\label{summary}

We have demonstrated the wide ranging applicability of the QMC model in different areas of physics.
The novelty of this approach lies in modeling the modification of the internal structure of the nucleon resulting 
from the large scalar mean fields in a nuclear medium. In turn, this leads to a \textit{microscopically derived} 
density dependence of the effective forces between hadrons in-medium, which is not limited to low densities. 
The calibrated few parameters are a single, universally applicable set,  in contrast to the larger parameter sets 
commonly used phenomenologically, which are often locally fine-tuned and lacking universality.

Finite nuclear systems are described at a similar level of accuracy as the currently more widely used Skyrme parametrizations.
Although the low-density QMC based force could be qualified as being of  a  'Skyrme-type' force and a relation between the 
QMC and Skyrme parameters can be found on a certain level of the model, there are important differences. 
The various terms in the QMC density functional all have a well identified physical origin. In particular, 
the non-trivial density dependence is directly related to the response of the nucleon quark structure  to the medium. 
The spin-orbit interaction has a rather subtle structure with the precession part, which is the same for any point-like 
spin 1/2 particle and the magnetic part, which has the right magnitude only if the particle has the correct magnetic 
moment, a constraint that is well satisfied in the bag model. The unambiguous prediction of the suppression of the 
spin-orbit force for $\Lambda$-hypernuclei is particularly notable. The derivation of the spin-orbit  interaction shows clearly 
that quark structure and relativistic effects are vital for low density nuclear physics. An important development of the 
QMC model was the formal derivation in terms of a mean field plus fluctuations. In the Hartree-Fock approximation 
the fluctuation term just generates the exchange terms but it will also play a role for excited states, 
an aspect which has not yet been studied. 

The relativistic version of the QMC model has been frequently compared to the performance of many variants of the 
RMF non-linear Walecka model in nuclear matter and neutron stars by variety of authors (see 
references in Section~\ref{nuclearmatter}). The predictions of the two models seem to be similar, but no conclusion 
has been drawn which model is closer to reality. One of the reasons for this situation is that there are no 
accurate data at present which would show a clear preference for either model. From theoretical point of view, 
there is however a fundamental difference in that in RMF models the Hartree potential depends on parameters 
which have to be fitted to experiment, while these can be calculated from the model in the QMC framework, 
without a need for extra parameters.

The latest version of the QMC model for finite nuclei, presented in Section~\ref{qmc}, is expected to be further 
developed and applied in several directions. There has been already work started through the implementation 
of the model in the HFODD framework~\cite{Schunck2017}, which will allow calculation of ground state 
properties of odd-A and odd-odd nuclei in a Hartree-Fock-Bogoliubov framework. This work is in progress in 
collaboration with the University of Warsaw. The HFODD model will allow also other investigation of the 
QMC model, including RPA, broken symmetry restoration and angular momentum projection.

It would be desirable to develop the relativistic version of the model in two major directions, namely an 
extension to finite temperatures and an investigation of the presence of quarks, together with the hadrons, 
in high density matter. The former will allow the study of proto-neutron stars and the latter may yield predictions for 
the interplay between the quark and hadron degrees of freedom in the super-saturated matter in compact objects, 
as well as the influence on their properties, such a maximum mass and radius of neutron stars.  Keeping in mind 
that we are dealing with \textit{cold} quark matter which is different from the matter made of deconfined quarks 
at high temperature and low or zero density. This will be a new area of investigation for high density matter.

Finally, one must not forget that central to the QMC model approach is the hypothesis that the structure of the objects 
bound in shell-model orbitals differs from that of free protons and neutrons and it is imperative that one strive 
to establish evidence that this is in fact the case. In Section~\ref{beyond} we reviewed recent developments regarding 
the EMC effect and the Coulomb sum rule, both of which hold the promise of new experimental results in the near future.

\section{Acknowledgement}
J. R. S. and P. A. M. G. acknowledge with pleasure support and hospitality of CSSM at the University of Adelaide 
during visits in the course of this work. It is a pleasure to acknowledge the technical support of R. Adorjan-Rogers 
during the intense computational phase of the project. The input of Kay Marie Martinez is gratefully acknowledged 
in connection with the results of QMC-I-$\pi$ and QMC-II. This work was supported by the University of Adelaide and 
the Australian Research Council, through funding to the ARC Centre of Excellence in Particle Physics at the Terascale 
(CE110001104) and Discovery Project DP150103101.

\section{References}

\bibliography{Biblio_Pierre_18}

\section{Appendix}

In this work we limit our considerations to the spin 1/2 SU(3) octet
$(N,\Lambda,\Sigma,\Xi)$ and therefore a flavor state can be specified
as $|f\rangle=|tms\rangle$ with $t,m$ the isospin and its projection and $s$
the strangeness, see Table~\ref{cap:QMC-Octet}. 

n the quark flavor space $(u,d,s)$ we have the matrices
\begin{equation}
\vec{I}=\left[\begin{array}{cc}
[\vec{\tau}/2] & 0\\
0 & 0
\end{array}\right],\,\,\Pi=\left[\begin{array}{ccc}
1 & 0 & 0\\
0 & 1 & 0\\
0 & 0 & 0
\end{array}\right],\,\, S=\left[\begin{array}{ccc}
0 & 0 & 0\\
0 & 0 & 0\\
0 & 0 & -1
\end{array}\right]\label{eq:B1}
\end{equation}
with $(\tau_{\alpha},\alpha=1,2,3)$ the $(2\times2)$ Pauli matrices
acting in the $(u,d)$ space.
 The strangeness $S$ is related to the
hypercharge by $S=Y+2/3$. We have introduced the matrix $\Pi=1+S$
which projects on the $(u,d)$ space because it occurs frequently.
Here are the matrix elements we need for the calculation:
I

\subsection{Projector}

Obviously
\begin{equation}
\langle tms|\sum_{q=1,3}\Pi(q)|t'm's'\rangle=\delta(tt')\delta(mm')\delta(ss')(3+s)\label{eq:B2}
\end{equation}
\begin{table}
\centering{}\caption{\label{cap:QMC-Octet}Quantum numbers of the octet members.}
\begin{tabular}{|c|c|c|c|c|c|c|c|c|}
\hline 
$b=$ & $p$ & $n$ & $\Lambda$ & $\Sigma^{-}$ & $\Sigma^{0}$ & $\Sigma^{+}$ & $\Xi^{-}$ & $\Xi^{0}$\tabularnewline
\hline 
\hline 
$t$ & $\frac{1}{2}$ & $\frac{1}{2}$ & 0 & 1 & 1 & 1 & $\frac{1}{2}$ & $\frac{1}{2}$\tabularnewline
\hline 
$m$ & $\frac{1}{2}$ & $-\frac{1}{2}$ & 0 & -1 & 0 & 1 & $-\frac{1}{2}$ & $\frac{1}{2}$\tabularnewline
\hline 
$s$ & 0 & 0 & -1 & -1 & -1 & -1 & -2 & -2\tabularnewline
\hline 
\end{tabular}
\end{table}
\subsection{Isospin:}

The matrix element 
\[
\langle tms|\sum_{q=1,3}\vec{I}(q)|t'm's'\rangle
\]
is independent of $(s,s')$ and is diagonal in $(tt').$ We note
\begin{equation}
\vec{I}_{mm'}^{t}=\langle tm|\sum_{q=1,3}\vec{I}(q)|tm'\rangle\label{eq:B3}
\end{equation}
We have $\vec{I}^{0}=0,\,\,\vec{I}_{mm'}^{1/2}=[\vec{\tau}/2]_{mm'}$
and 
\begin{equation}
\vec{I}^{1}.\vec{e}(3)=\left[\begin{array}{ccc}
1 & 0 & 0\\
0 & 0 & 0\\
0 & 0 & -1
\end{array}\right],\,\,\vec{I}^{1}.\vec{e}(1)=\frac{1}{\sqrt{2}}\left[\begin{array}{ccc}
0 & 1 & 0\\
1 & 0 & 1\\
0 & 1 & 0
\end{array}\right],\,\,\vec{I}^{1}.\vec{e}(2)=\frac{1}{\sqrt{2}}\left[\begin{array}{ccc}
0 & -i & 0\\
i & 0 & -i\\
0 & i & 0
\end{array}\right]\label{eq:B3.1}
\end{equation}
The general expression is 
\begin{equation}
\vec{I}_{mm'}^{t}=t\sqrt{6}\sum_{\mu}\vec{e}^{*}(\mu)(-)^{t-m}\left(\begin{array}{ccc}
t & 1 & t\\
-m & \mu & m'
\end{array}\right),\,\,\,\, t=0,\frac{1}{2},1\label{eq:B4}
\end{equation}
with $\vec{e}(0,\pm)$ the standard unit vectors. One has the relations
(no summation over $m,m'$) 
\begin{eqnarray}
\vec{I}_{mm}^{t} & = & \vec{e}(3)m,\label{eq:B5}\\
\vec{I}_{mm'}^{t}.\vec{I}_{m'm}^{t} & = & \delta_{mm'}m^{2}+t(\delta_{m,m'+1}+\delta_{m',m+1})\label{eq:B6}
\end{eqnarray}

\subsection{Projector-spin}

One has

\begin{equation}
\langle f,\sigma|\sum_{q=1,3}\Pi(q)\vec{\sigma}(q)|f',\sigma'\rangle=C(f)\delta(ff')\vec{\sigma}_{\sigma\sigma'}\label{eq:B7}
\end{equation}
and using the spin flavor wave functions one gets
\begin{equation}
C(p,n)=1,\,\, C(\Lambda)=0,\,\, C(\Sigma's)=\frac{4}{3},\,\, C(\Xi's)=-\frac{1}{3}.\label{eq:B8}
\end{equation}

\subsection{Gamow - Teller}

We define the Gamow -Teller operator acting on the baryon of spin-flavor
$\sigma,b=tms$ by the matrix element
\[
\langle\sigma,b|\vec{G}_{T}^{\mu}|\sigma',b'\rangle=\langle\sigma,b|\sum_{q=1,3}\vec{\sigma}(q)\vec{I}^{\mu}(q)|\sigma',b'\rangle
\]
Using the Wigner Eckart theorem we can write
\begin{equation}
\langle\sigma,b|\vec{G}_{T}^{\mu}|\sigma',b'\rangle=\delta(ss')\vec{\sigma}_{\sigma\sigma'}(-)^{t-m}\left(\begin{array}{ccc}
t & 1 & t\\
-m & \mu & m'
\end{array}\right)\langle t,s||\vec{\sigma}\vec{I}||t',s\rangle\label{eq:B9}
\end{equation}
and the explicit calculation gives
\begin{eqnarray}
 &  & \langle\frac{1}{2},0||\vec{\sigma}\vec{I}||\frac{1}{2},0\rangle=\frac{5}{\sqrt{6}},\,\,\langle1,-1||\vec{\sigma}\vec{I}||1,-1\rangle=\frac{2\sqrt{2}}{\sqrt{3}},\,\,\langle\frac{1}{2},-2||\vec{\sigma}\vec{I}||\frac{1}{2},-2\rangle=-\frac{1}{\sqrt{6}}\nonumber \\
 &  & \langle0,-1||\vec{\sigma}\vec{I}||0,-1\rangle=-\langle1,-1||\vec{\sigma}\vec{I}||0,-1\rangle=1.\label{eq:B10}
\end{eqnarray}

\end{document}